\documentclass[journal]{IEEEtran}

\usepackage{cite}
\usepackage{amsmath}
\usepackage{amsfonts}
\usepackage{amssymb}
\usepackage{makeidx}
\usepackage{multicol}
\usepackage{mathrsfs}
\usepackage{amsthm}
\usepackage{url}
\usepackage{color}
\usepackage{multirow}

\ifCLASSINFOpdf
   \usepackage[pdftex]{graphicx}
\else
\fi

\newtheorem{theorem}{Theorem}[]
\newtheorem{lemma}[theorem]{Lemma}
\newtheorem{proposition}[theorem]{Proposition}
\newtheorem{corollary}[theorem]{Corollary}
\newtheorem{definition}[theorem]{Definition}

\DeclareMathOperator{\trace}{trace}
\DeclareMathOperator{\diag}{diag}
\DeclareMathOperator{\Beta}{\boldsymbol\beta}
\DeclareMathOperator{\sinrmmsePil}{\widehat{\mathsf{SINR}}}
\DeclareMathOperator{\sinrmmsePer}{\mathsf{SINR}^*}
\DeclareMathOperator{\sinrmatPil}{\overline{\mathsf{SINR}}}
\DeclareMathOperator{\filmmsePil}{\mathbf{\hat{c}}}
\DeclareMathOperator{\filmmsePer}{\mathbf{c}^*}
\DeclareMathOperator{\mseInt}{\mathcal{C}(\alpha)}
\DeclareMathOperator{\SignalPwr}{ \tilde{P}_{\mathsf{signal} }}
\DeclareMathOperator{\PilotPwr}{ \tilde{P}_{\mathsf{contam} }}
\DeclareMathOperator{\InterPwr}{ \tilde{P}_{\mathsf{inter} }}

\newcounter{MYtempeqncnt}

\allowdisplaybreaks
\begin{document}
\title{ Uplink  Linear Receivers for Multi-cell Multiuser MIMO with Pilot Contamination: Large System Analysis}

\author{    \authorblockN{Narayanan Krishnan, Roy D. Yates, Narayan B. Mandayam \\}
\authorblockA{WINLAB; Rutgers, The State University of New Jersey\\
E-mail: narayank, ryates, narayan @winlab.rutgers.edu}
    }

\maketitle \thispagestyle{empty}
\begin{abstract}
Base stations with a large number of transmit antennas have the potential to serve a large number of users at high rates. However, the receiver processing in the uplink relies on channel estimates which are known to suffer from pilot interference. In this work, making use of the similarity of the uplink received signal in CDMA with that of a multi-cell multi-antenna system, we perform a large system analysis when the receiver employs an MMSE filter with a pilot contaminated estimate. We assume a Rayleigh fading channel with different received powers from users. We find the asymptotic Signal to Interference plus Noise Ratio (SINR) as the number of antennas and number of users per base station grow large while maintaining a fixed ratio. Through the SINR expression we explore the scenario where the number of users being served are comparable to the number of antennas at the base station. The SINR explicitly captures the effect of pilot contamination and is found to be the same as that employing a matched filter with a pilot contaminated estimate. We also find the exact expression for the interference suppression obtained using an MMSE filter which is an important factor when there are significant number of users in the system as compared to the number of antennas. In a typical set up, in terms of the five percentile SINR, the MMSE filter is shown to provide significant gains over matched filtering and is within $5$~dB of MMSE filter with perfect channel estimate. Simulation results for achievable rates are close to large system limits for even a $10$-antenna base station with $3$ or more users per cell.

\end{abstract}


\IEEEpeerreviewmaketitle
\section{Introduction}
\label{sec:Introduction}

 Cellular systems with large number of base station antennas have been found to be advantageous in mitigating the fading effects of the channel \cite{Marzetta2011} while increasing system capacity. 
 It is shown in \cite{Marzetta2011} that in an infinite antenna regime, and in a bandwidth of $20$~MHz,  a time division duplexing system has the potential to serve $40$ single antenna users with an average throughput of $17$~Mbps per user. However, any advantages offered by multiple antennas at the base station can be utilized only by gaining the channel knowledge between the base station and all the users. This requires training data to be sent from the users. In a typical system the time-frequency resources are divided into Physical Resource Blocks (PRBs) of coherence-time coherence-bandwidth product. For each user, it is necessary and sufficient to estimate the channel in every PRB. Thus, some resources (time slots or equivalently frequency channels) are used for channel estimation and the rest are used for transmission in uplink or downlink. However, in \cite{Marzetta2006}, it has been shown that the number of pilot symbols required is proportional to the total number of users in the system. Hence, as the system scales with the number of users, the dedicated training symbols may take up a significant portion of the PRB.  As this is undesirable, only a part of the coherence time is utilized to learn the channel. In this case, the pilot sequences in different cells overlap over time-frequency resources and, as a consequence, the channel estimates are corrupted. This \emph{pilot interference} is found to be a limiting factor as we increase the number of antennas \cite{Jubin}. 

It is shown in \cite{Marzetta2011} that in the limit of large number of antennas, the SINR using a matched filter receiver is limited by interference power due to pilot contamination. While the result assumes a regime with finite number of users, we can also envision a regime where the number of users may be comparable to the number of antennas such as a system with $50$ antenna base stations serving $50$ users simultaneously. 
In this work, we do a large system analysis of uplink multi-cell, multi-antenna system when the receiver employs an MMSE filter to decode the received signal. We investigate the SINR in a regime where the number of users per cell  is comparable to the number of antennas at the base station. The MMSE filter which is designed to maximize the SINR is evaluated when we have a pilot corrupted channel estimate.
We let the number of antennas and the number of users per base station grow large simultaneously while maintaining a fixed users/antennas ratio $\alpha$ and observe the SINR for the above two cases as a function of $\alpha$. To do so we make use of the similarity of the uplink received signal in a MIMO system to that of the received signal in a CDMA system \cite{TseHanly1999}.

 Much of the research in large MIMO systems with Rayleigh channel can be borrowed from the considerable literature for CDMA systems. The channel vector with i.i.d entries for the large MIMO system is analogous to the signature sequence in a CDMA system so that antennas contribute to the processing gain. For example, the uplink analysis of an  asymptotic regime \cite{TseHanly1999} with both users and signature sequences tending to infinity translates directly to results in a large MIMO system when signatures are replaced by antennas. In both systems it is observed that the asymptotic analysis is a good approximation for practical number of antennas (signatures) and users. While in a CDMA system we assume that the signature sequences are known, there are practical limitations in learning a mobile radio multi-antenna channel (antenna signatures) in a multi-cellular system, as shown in \cite{Marzetta2011}. In this work we explore this limitation when users simultaneously estimate the channel and the estimates are subject to  pilot contamination. We focus our results in the regime $\alpha > 0.1$ as opposed to recent works such as \cite{Marzetta2011,Marzetta:2013,Hoydis:2013,Ngo:2011} which are found to be approximately in the regime of $ 0 \leq \alpha < 0.1$. Further, we compare the results of the asymptotic SINR expression so obtained with that of the performance of the matched filter. 

\subsection{Related Work}
A similar large system analysis in the context of a Network-MIMO architecture was presented in \cite{Caire:2012}. The authors concluded that high spectral efficiencies can be realized even with $50$ antennas in their architecture, paralleling the existing literature results in CDMA systems. In \cite{Ngo:2011}, the results obtained suggest to scale the transmission power by the square root of the number of base station antennas,  as opposed to scaling by the number of antennas. This  assumes that the transmission power during training and data are same. In general we take the  approach in \cite{Babak:2003} where the transmission power can be different for the training and data symbols during a coherence time. Joint channel estimation and multi-user detection was considered in \cite{Takeyuchi:2012} for a single cell multi-user MIMO DS-CDMA systems. For fixed  number of  transmit antennas per user and receive antennas per base station, the authors  employ replica method for CDMA large system analysis in order to obtain lower bounds on achievable rate for different feedback based receiver strategies. Mathematically, our model can be viewed as an extension of the linear receiver scheme considered in their model to a multi-cell scenario. 

In the MU-MIMO literature the asymptotic SINR is also called  the ``deterministic equivalent" of the SINR. Recent work in \cite{JZHANG:2013} finds the deterministic equivalent of SINR with distributed sets of correlated antennas in the uplink. Authors in \cite{Hoydis:2013}, have done a considerable work in providing the deterministic equivalent for beamforming/maximum ratio combining and regularized zero forcing/MMSE in the downlink/uplink with a generalized channel model taking into account the effect of pilot contaminated channel estimate. They find the number of antennas required to match a fixed percentage of the rate of an infinite antenna regime. Also, the number of extra of antennas required for the matched filter to equal the rate obtained out of the MMSE filter is shown, implicitly showing the interference suppression capability of MMSE filter. We derive in our work the exact amount by which the MMSE filter suppresses the interference for a Rayleigh fading channel and provide some fresh engineering insights in the regime with $\alpha > 0.1$. Additionally, we derive those results  under a stochastic rather than an deterministic received power model as presented in \cite[section 4]{Hoydis:2013}. A summary of the results is given in section \ref{sec:contri}.

There have been significant attempts to mitigate pilot contamination in the recent works in ``Massive MIMO" systems which are, however, specific to the case when antennas far exceed the number of users served i.e. $\alpha < 0.1$. In \cite{Marzetta:2013}, time shifted pilot schemes were introduced to reduce pilot contamination. There, simultaneous transmission of pilots was avoided by scheduling only a subset of base stations to transmit uplink pilots. Simultaneously, other base stations transmit in the downlink to their users and it is shown that the interference created by these downlink transmission can be cancelled with a large number of antennas at the base station estimating the channel. Pilot contamination is now restricted to base stations in a group that simultaneously transmit uplink pilots. However, this requires that the number of antennas far exceed the number of users. In their recent work, authors in \cite{Müller2013,Muller:CTW:2013} show that pilot contamination can be avoided using subspace based channel estimation techniques. They show that the eigenvalues corresponding to the other-cell interference subspace can be separated from the in-cell users in a regime where $\alpha$ is below a threshold. Their analyses assume an ideal power controlled situation with strict user scheduling and antennas far exceeding the number of users. By contrast, we examine the operating regime in which $ 0.1 < \alpha$ and determine the effect of pilot contamination on interference and interference suppression capability of MMSE receiver. Also, even with power control, pilot contamination is prevalent when linear MMSE channel estimation is employed.

\subsection{Contributions of our Work}
\label{sec:contri}
 We develop a large system asymptotic expression for the SINR in a Rayleigh fading environment when using an matched filter and MMSE filter with a pilot contaminated channel estimate for a arbitrary user in the system. The SINR expression is dependent on the number of users and the number of antennas only through their ratio $\alpha$. If $\SignalPwr$, $\PilotPwr$, $\InterPwr (\mathbf{c})$ and $\sigma^2$ represent the signal power, interference power due to pilot contaminated estimate, the filter dependent interference power for a linear filter $\mathbf{c}$ and the noise variance respectively, we show that the expression for asymptotic SINR for both matched filter and MMSE filter can be generalized to,
 \begin{eqnarray}
 \mathsf{SINR}(\mathbf{c}) = \frac{\SignalPwr}{ \sigma^2 + \PilotPwr + \alpha \InterPwr (\mathbf{c})}.
 \end{eqnarray}
 If $\hat{\mathbf{c}}^{\mathsf{MF}}$ and $\hat{\mathbf{c}}^{\mathsf{MMSE}}$ denote the matched filter and the MMSE filter with a pilot contaminated estimate, then the following result summarizes our contribution:
\begin{itemize}
\item From the expression for SINR it is derived that the total interference is the sum of two terms. The first  term given by $\PilotPwr$ is due to employing pilot contaminated channel estimate and the second is the filter dependent interference term $\alpha \InterPwr (\mathbf{c})$ due to comparable $K$ and $M$. Also, the the filter dependent term contributes to interference only when $\alpha \neq 0$.
\item We show that $\InterPwr(\hat{\mathbf{c}}^{\mathsf{MF}}) - \InterPwr(\hat{\mathbf{c}}^{\mathsf{MMSE}}) = \mseInt \geq 0$, where $\mseInt$ is called the interference suppression term. We find a closed form expression for $\mseInt$ showing the interference suppression capability of the MMSE filter when $\alpha >0$.
\item It is derived that $\PilotPwr (\hat{\mathbf{c}}^{\mathsf{MF}}) = \PilotPwr(\hat{\mathbf{c}}^{\mathsf{MMSE}}) = \PilotPwr$. The contribution of the pilot contaminated channel estimate to the pilot interference given by $\PilotPwr$ is same for both matched filter and MMSE filter. 
\item Although the filter $\hat{\mathbf{c}}^{\mathsf{MMSE}}$ depends on the pilot contaminated channel estimate of all the $K$ users in a cell to its base station, the $\PilotPwr$ is independent of $K$ or $M$ or $\alpha$ for all values of $\alpha>0$ and therefore is the same as what was found at $\alpha = 0$.
\end{itemize}
 As per the authors' knowledge, these contributions have not been reported yet in the literature.
 
We validate the theoretical results with simulations.  We show that even a system with $50$ base station antennas each serving some number of users sufficiently qualifies for the term large system as the users' SINRs are close to the asymptotic limit. Simulation results for achievable rates are close to theory for even a $10$-antenna base station with $3$ or more users per cell. The following summarizes the key contributions through simulation:
\begin{itemize}
\item The theoretical results are derived assuming that the same set of in-cell orthogonal training signals are repeated across the cells. However, we also show through simulations that in the case of independently generated but non-orthogonal training signals(with orthogonal in-cell training), the resulting SINR performance is close to the asymptotic limit. 
\item In an example seven cell set up, the MMSE filter performs the best in the absence of pilot contamination. We also show an intermediate regime where the MMSE filter with pilot contamination obtains around $7$~dB gain over the matched filter with a pilot contaminated estimate. 
\item In terms of the five percentile SINR, the MMSE receiver is shown to provide significant gains over matched filtering. Also in most of the operating points $\alpha$, the performance of the MMSE receiver with pilot estimate is within $5$~dB of the MMSE filter with perfect estimate. 
\item We also show that the achievable rates are within a $1$~bit/symbol of the MMSE filter with perfect estimate when the number of users are comparable to the number of antennas.
\end{itemize}

\section{System Model}
\label{sec:system_model}
We consider a system similar to that in \cite{Marzetta2011} with $B$ non-cooperating base stations and $K$ users per base station. We assume that all $KB$ users in the system are allocated the same time-frequency resource by a scheduler. Also, each base station is equipped with $M$ antennas. The channel vector representing the small scale fading between  user $k$ in cell $j$ and the antennas in base station $l$ is given by a $M \times 1$ vector $\mathbf{h}^{(l)}_{jk}$. The entries of $\mathbf{h}^{(l)}_{jk}$ are assumed to be independent zero mean i.i.d Gaussian random variables with variance $1/M$ corresponding to the scaling of transmit power by the number of receiver antennas at the base station. This corresponds to an ideal and favourable propagation medium with rich scattering. A large scale fading coefficient, which represents the power attenuation due to distance and effects of shadowing between base station $l$ and $k$th user in $j$th cell is given by $\beta^{(l)}_{jk}$. We assume that $\beta^{(l)}_{jk} < 1$ as we do not expect the received power to be greater than what is transmitted.  This is constant across the antennas of the cell $l$. Accordingly, overall channel vector is given by $\sqrt{\beta^{(l)}_{jk}}\mathbf{h}^{(l)}_{jk} $. 

\subsection{Uplink Transmission}
\label{subsec:UplinkTransm}
 We assume that all users' transmission are perfectly synchronized.  Also, while a user's transmission is intended for its own base station, other base stations also hear the transmission. Defining $q_{jk}$ as the symbol transmitted by user $k$ in cell $j$,
$\mathbf{w}^{(l)}$  as the $M\times 1$ noise vector with zero mean circularly symmetric Gaussian entries such that $\mathbb{E}[\mathbf{w}^{(l)}{\mathbf{w}^{(l)}}^H] = \sigma^2 \mathbf{I}$, the received signal at base station $l$ is given by,
\begin{eqnarray}
\mathbf{y}^{(l)} &=&  \sum^B_{j=1} \sum^K_{k=1} \sqrt{\beta^{(l)}_{jk}} \mathbf{h}^{(l)}_{jk}q_{jk}   +  \mathbf{w}^{(l)} \label{eq:RcvdSgnlVectorForm}\text{.} 
\end{eqnarray}
Here, the Signal to Noise Ratio (SNR) of the received signal per receive dimension is given by $1/\sigma^2 M$. 
 In order to utilize the advantages offered by multiple antennas, the base station has to have an estimate of the channel to all users prior to detection of uplink signals.  In a system employing an OFDM physical layer with time-frequency resources, we can divide the resources into physical resource blocks (PRBs) contained in the coherence-time coherence-bandwidth product. Although the channel vector $\mathbf{h}^{(l)}_{jk}$ of  each user has to be relearned by the base station at the start of PRB, once learnt for a subcarrier it remains the same for all subcarriers within that PRB. Let the number of coherent symbols be given by $T_c$ and coherent subcarriers be $N_c$. We define a Resource Element(RE) to be a subcarrier at a symbol time. Therefore, if we fix the number of resource elements used for estimation to be $T$ such that $T \leq T_cN_c$, a total of $T$ user's channel can be learnt. This observation was noted in \cite{Marzetta2011}. We would like to point out that it is relevant here as the number of users that can be supported depends on total available coherent resource elements $N_cT_c$. Depending on its value the number of users per cell $K$ that could be supported can be comparable to $M$. We define the load on the system as $\alpha = K/M$ throughout this paper. For example, in a single cell set up even in very conservative scenarios of short coherence time with $T_c = 7$ symbols and frequency selective channel with $N_c = 14$, around $49$ users per cell can be supported at $\alpha = 0.5$ if half the resources are used for channel estimation. This is also illustrated in figure \ref{fig:PRB}. Therefore, it is worthwhile to investigate not only the $M \gg K$ scenario but also the case when $M$ and $K$ large and comparable. 
 
 \subsection{Limitations in gaining Channel Knowledge }
\begin{figure}[t]
    \begin{center}
      \includegraphics[width = 3.5in, clip = true, trim = 10 0 10 0]{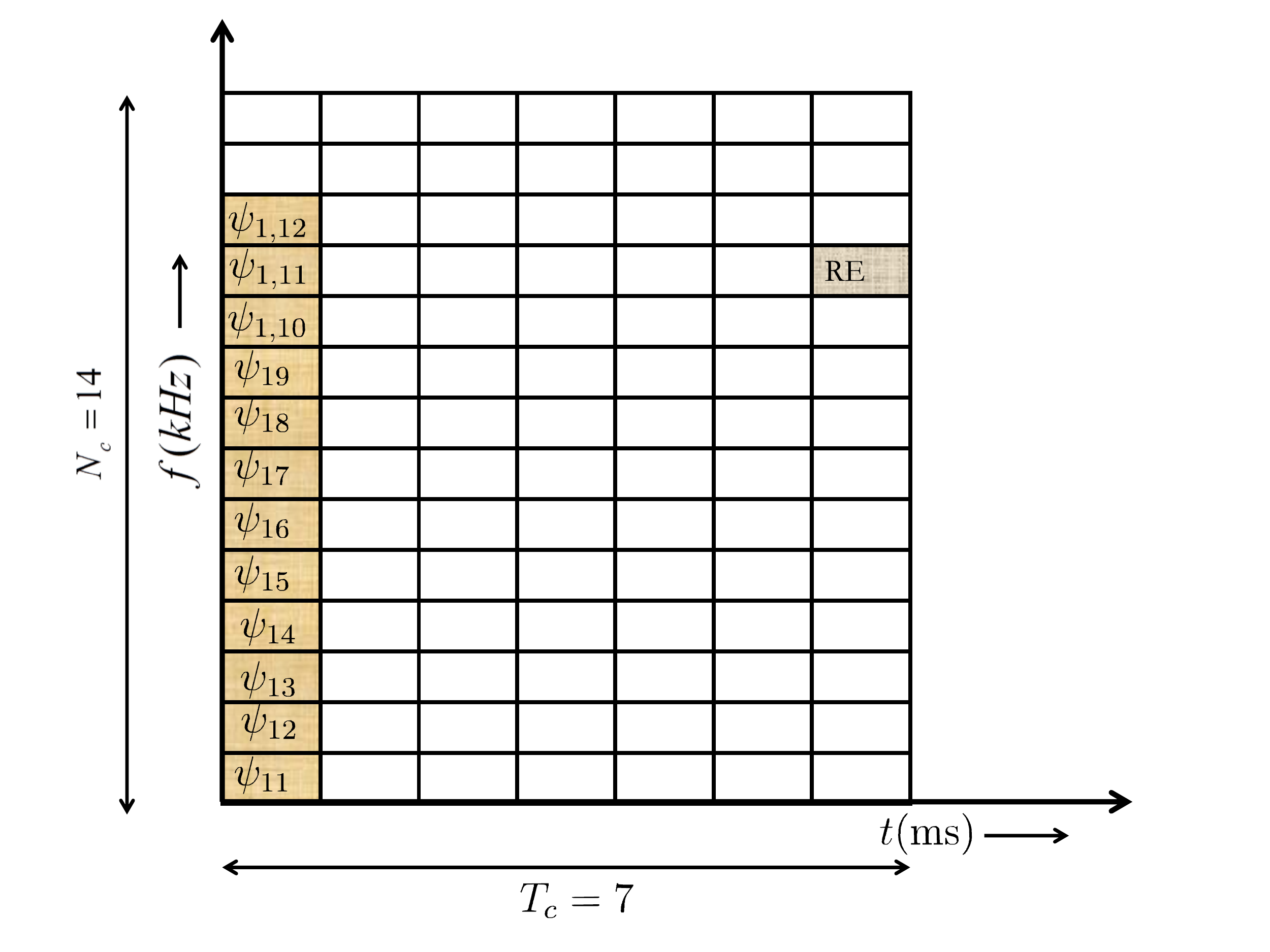}
      \caption{ The PRB is composed of $98$ orthogonal Resource Elements(RE) and $12$ RE for training. The first user in the first base station transmits pilots $\psi_{11}$ in the first RE and remain silent for the rest of training duration. Similarly, other users in the first cell transmit pilots during the RE allocated for its training. If orthogonal pilots are allocated for all the users in the system, a typical $7$ cell system $14$ users per cell would consume all the coherent RE without any time for data transmission. In our model we let the training symbols of other cell users $\psi_{jk}$ for all $j$ and $k$ simultaneously during $\psi_{1k}$ leading to pilot contamination problem. } 
      \label{fig:PRB}
    \end{center}
\end{figure}

During each coherence time, users in a cell spend some pilot symbol times in each PRB for channel estimation at the base station and then data transmission ensues until the end of the block. At base station $l$, the number of channel vectors $\mathbf{h}^{(l)}_{jk}$  that needs to be learnt is equal to the number of users in the system which is $KB$. In order to accomplish that, the number of pilots required must at least be $KB$ symbol times in order for the pilot sequences to be orthogonal across the users in the system. However, such a system will not be scalable as there exists some large $B$ for which the product $KB$ will occupy all the coherent resource elements. In the example illustrated in figure \ref{fig:PRB} a $7$ cell system with $14$ users per cell would end up using all the coherent resource elements if orthogonal channel training is provided for all the users in the system. This is clearly undesirable as pilot training is consuming a significant part of PRB.

In one of the approaches taken in \cite{Marzetta2011}, the base station is concerned with only knowing the channel to its own $K$ users and spends only $K$ resource elements for channel estimation instead of $KB$. Every base station similarly spends  $K$ resource elements for channel estimation for its $K$ users.  The pilot signals are processed and an MMSE based channel estimate of the channel is formed. MMSE channel estimation is the commonly employed in multiuser MIMO systems \cite{Babak:2003, Dragan:2007, Jubin:2009}. Let $\Psi_{jk} \in \mathbb{C}^{K \times 1}$ denote the training sequence of user $k$ in cell $j$ of duration $K$ symbols. Also assume that the in-cell training sequences are orthogonal i.e $\Psi^H_{jk} \Psi_{jn} = 0$ if $k \neq n$ and $1$ otherwise. We assume that the training sequences across the cells are independently generated and hence in general $\Psi^H_{jk} \Psi_{in} \neq 0$ if $i \neq j$ and for all $k$ and $n$. We assume a constant average transmit SNR per symbol during a coherence time. If $\rho_d$ is the transmit SNR of data symbol and $\rho_{\mathsf{avg}}$ the average transmit SNR per symbol then with $K$ resource elements for channel estimation we have,
\begin{eqnarray}
\rho_{\mathsf{avg}} = \frac{\rho_p K + \rho_d(T_cN_c - K)}{T_c N_c}.
\end{eqnarray} 
If the transmit power of data symbols is scaled by the number of antennas then $\rho_d = 1/\sigma^2 M$. Therefore, $\rho_d \rightarrow 0$ for large $M$ and $\rho_p \approx \rho_{\mathsf{avg}} T_c N_c/K$. Therefore, in the constant average transmit SNR per symbol model the pilot transmission is not scaled by $M$ which is the number of antennas at the base station. With $\mathbf{N} \in \mathbb{C}^{M \times K}$  denoting the additive complex Gaussian noise matrix with i.i.d entries with variance $1/M$, and $\rho_p$ denoting the SNR during pilot transmission, the received signal at base station $l$ across the $K$ training resources is given by,
\begin{eqnarray}
\label{eq:PilotSignals}
\mathbf{Y}^{(l)} =   \sum^B_{j=1} \sum^K_{k=1} \sqrt{\beta^{(l)}_{jk}} \mathbf{h}^{(l)}_{jk} \Psi^H_{jk} + \frac{\mathbf{N}^{(l)}}{\sqrt{\rho_p}}.
\end{eqnarray}
 Since the transmission power of pilot symbols are not scaled by the number of antennas we assume that noise variance $1/M$ for the entries of the $\mathbf{N}^{(l)}$. This is required because we have defined the entries of $\mathbf{h}^{(l)}_{jk}$ to be complex i.i.d with variance $1/M$.

Without loss of generality we assume that the receiving base station is indexed $l=1$ throughout this paper. Consequently, to simplify the exposition we drop the superscript $(.)^{(l)}$ from terms in equation~(\ref{eq:PilotSignals}) and denote $\mathbf{Y}^{(1)} \triangleq \mathbf{Y}$, $\mathbf{N}^{(1)} \triangleq \mathbf{N}$, $\beta^{(1)}_{jk} \triangleq \beta_{jk}$, $\mathbf{h}^{(1)}_{jk} \triangleq \mathbf{h}_{jk}$. Hence,  $\beta_{jk}$ and $\mathbf{h}_{jk}$ represent the large scale fading and the small scale channel vector  between the user $k$ in cell $j$ to cell $1$. Also in equation~(\ref{eq:RcvdSgnlVectorForm}), if $l=1$ then $\mathbf{y}^{(1)} \triangleq \mathbf{y}$ and $\mathbf{w}^{(1)} \triangleq \mathbf{w}$. The MMSE channel estimate for user $k$ in the first cell is then given by,
\begin{eqnarray}
\label{eq:trainingBasedEst}
\hat{\mathbf{h}}_{1k} = \mathbf{Y} \left( \frac{\mathbf{I}}{\rho_p} + \sum^B_{j=1} \sum^K_{k=1} \beta_{jk} \Psi_{jk} \Psi^H_{jk}\right)^{-1} \Psi_{1k} \sqrt{\beta}_{1k}.
\end{eqnarray}

Here, $\hat{\mathbf{h}}_{1k} \left( \triangleq \hat{\mathbf{h}}^{(1)}_{1k} \right)$ is the channel estimate of the user $k$ in cell $1$ to the base station $1$. Although it is not common in practice, we assume as in \cite{Marzetta2011,Jubin:2009,Jubin} that the in-cell pilots are repeated across the cells, in order to get some analytic insight; this implies that $\Psi_{jk} = \Psi_{ik}$ for all $k$. The MMSE channel estimate \cite{HVPoor} with pilot contamination when the in-cell orthogonal pilots \cite{Jubin:2009} are repeated across the cells is given by,
\begin{eqnarray}
 \mathbf{\hat{h}}_{1k} = \frac{\sqrt{\beta_{1k}}}{\beta^{(k)} + \frac{1}{\rho_p}} \left( \sum^B_{j=1}\sqrt{\beta_{jk}}\mathbf{h}_{jk} + \frac{\mathbf{N}^{} \Psi_{1k}}{\sqrt{\rho_p}} \right)
 \label{eq:PilotContChannelEst0}
\end{eqnarray}
where, $\beta^{(k)} = \sum^B_{j=1} \beta_{jk}$ and if $\mathbf{h}_{1k} = \hat{\mathbf{h}}_{1k} + \tilde{\mathbf{h}}_{1k}$, then $\tilde{\mathbf{h}}_{1k}$ is zero mean with covariance 
\begin{eqnarray}
\mathbb{E} \left[ \tilde{\mathbf{h}}_{1k}  \tilde{\mathbf{h}}^H_{1k}\right]  = \frac{1}{M}\left( \frac{\sum^B_{j=2} \beta_{jk} + 1/\rho_p}{\beta^{(k)} + 1/\rho_p} \right)\mathbf{I}. \nonumber
\end{eqnarray}
This estimate is used to design linear detectors to filter the received signal. Later we show that even with actual training given by equation~(\ref{eq:trainingBasedEst}) the SINRs are very close to when in-cell pilots are reused across the cells. In a power controlled system with $\beta_{1k} = 1$, for all $k = 1, \dots K$, if the target average SNR is $\rho_{\mathsf{avg}} = 20$~dB, then in the example in figure~\ref{fig:PRB}, the SNR of pilot transmission is given by $\rho_p \approx \rho_{\mathsf{avg}} T_c = 28$~dB. As the above example shows, even with a very conservative coherence time $T_c$, the SNR of pilot transmission is high enough such that $\sum^B_{j=2} \beta_{jk} \gg 1/ \rho_p$ unless all the other cell received powers of interferers contributing to pilot contamination are approximately $28$~dB below the in-cell user. As we will see in simulations, having other-cell received powers $30$~dB below in-cell corresponds to a situation in which pilot contamination is insignificant. In general since $\rho_p \approx  \rho_{\mathsf{avg}}T_cN_c/K$, we can always design a system based on fixing a percentage of resource elements in a PRB for training such that noise in the channel estimate is not the significant contributor. Therefore, assuming we have high enough pilot power, we ignore the additive noise affecting the channel estimation in order to focus our results on the pilot contamination problem. Therefore, with $\rho_p \rightarrow \infty$ the channel estimate is given by,
\begin{eqnarray}
 \mathbf{\hat{h}}_{1k} = \frac{\sqrt{\beta_{1k}}}{\beta^{(k)} } \sum^B_{j=1}\sqrt{\beta_{jk}}\mathbf{h}_{jk}.
 \label{eq:PilotContChannelEst}
\end{eqnarray}

\subsection{Linear Receivers}
We assume that the received signal is projected onto a linear filter $\mathbf{c}^{(l)}_{lk} \in \mathbb{C}^{M \times 1}$ for the $k^{th}$ user in the $l^{th}$ cell. Since the SINR analysis is identical for all users in the system we focus only on user $k =1$ in base station indexed $l = 1$. We also drop the superscript $(.)^{(1)}$ for notational convenience. Consequently, if coherent detection is employed then  $\mathbf{c}_{} = \mathbf{\hat{h}}_{11}$.  Alternatively, using the channel estimates for all users of the first cell, the MMSE filter for user $1$ in the cell $1$  is defined as $\text{arg}\min_{\mathbf{c}} \mathbb{E}\left[|q_{11} - \mathbf{c}^H \mathbf{y}|^2|\mathbf{\hat{h}}_{1k} \forall k\right]$. Defining
\begin{eqnarray}
\mathbf{z} = \sum^B_{j=2}\sum^K_{k=1}\sqrt{\beta_{jk}}\mathbf{h}_{jk}q_{jk}, 
\end{eqnarray}
which represents the other cell interference and $\mathbf{h}_{1k}=\mathbf{\hat{h}}_{1k} + \mathbf{\tilde{h}}_{1k}$ the received signal can be rewritten as,
\begin{eqnarray}
\mathbf{y} &=&  \sum^K_{k=1} \sqrt{\beta_{1k}} \mathbf{\hat{h}}_{1k} q_{1k} + \sum^K_{k=1} \sqrt{\beta_{1k}} \text{,} \mathbf{\tilde{h}}_{1k} q_{1k} + \mathbf{z} + \mathbf{w}
\end{eqnarray}
where, $\mathbf{\tilde{h}}_{1k}$ is the result of pilot contamination.
The MMSE filter is then given by the expression,
\begin{eqnarray}
\filmmsePil &=& \left( \mathbf{E}[\mathbf{y}\mathbf{y}^H|\mathbf{\hat{h}}_{1k}\forall k] \right)^{-1}\mathbf{E}[\mathbf{y}q^*_{11}| \mathbf{\hat{h}}_{1k} \forall k] \nonumber \\
\label{eq:MmseFilterPilot}
&=& \mathbf{S}^{-1} \sqrt{\beta_{11}} \mathbf{\hat{h}}_{11} ,
\end{eqnarray}
where,
\begin{eqnarray}
\label{eq:MmseFilterMatrix}
\mathbf{S} 
=\left( \sum^K_{k=2} \beta_{1k} \mathbf{\hat{h}}_{1k} \mathbf{\hat{h}}^H_{1k}  + (\theta_1 + \theta_2 + \sigma^2) \mathbf{I}\right), 
\end{eqnarray}
and,
\begin{eqnarray}
\label{eq:MmseFilterInter}
\theta_1 \mathbf{I} &=& \mathbb{E} \left[\mathbf{z}\mathbf{z}^H \right]  =  \sum^B_{j=2} \left[\frac{1}{M}\sum^K_{k=1} \beta_{jk}\right] \mathbf{I} ,  \\
\label{eq:MmseFilterEstErrInter}
\theta_2\mathbf{I} &=& \sum^K_{k=1} \beta_{1k}\mathbb{E}[ \mathbf{\tilde{h}}_{1k} \mathbf{\tilde{h}}^H_{1k}]  \nonumber \\
&=& \sum^B_{j=2} \left[\frac{1}{M}\sum^K_{k=1} \beta_{jk}\left( \frac{\beta_{1k}}{\beta^{(k)}}\right) \right]\mathbf{I} \text{.}
\end{eqnarray}
As seen from the expression for the filter in equation~(\ref{eq:MmseFilterPilot}), the lack of channel knowledge of other-cell users and only a partial channel knowledge of in-cell users shows up as effective noise terms $\theta_1$ and $\theta_2$ respectively. In order to obtain the expression we also use the properties of the MMSE estimate that $\mathbb{E}\left[\hat{\mathbf{h}} \tilde{\mathbf{h}}^H \right] = 0$. In an ideal situation, the channel estimation incurs no error and $\mathbf{\hat{h}}_{1k} = \mathbf{h}_{1k}$ for all $k$,  then 
\begin{eqnarray}
\label{eq:MmseFilterPer}
\filmmsePer =  \left( \sum^K_{k=1} \beta_{1k} \mathbf{h}_{1k} \mathbf{h}^H_{1k} + (\theta_1 + \sigma^2) \mathbf{I}\right)^{-1} \sqrt{\beta_{11}} \mathbf{h}_{11} 
\end{eqnarray}
This is an optimistic scenario which will serve as a benchmark for the performance of the MMSE filter with pilot contamination.
\section{MMSE Filter with Pilot Contaminated Estimate}
\label{sec:Main_Result}
After processing the received signal using the linear filter $\mathbf{c}$, let $P_{\mathsf{signal}}$, $P_{\mathsf{noise}}(\mathbf{c})$, $P_{\mathsf{contam}}(\mathbf{c})$, $P_{\mathsf{inter}}(\mathbf{c})$  denote the signal power, noise power, pilot interference power and interference power respectively as a function of the filter $\mathbf{c}$. It follows that,
\begin{eqnarray}
\label{eq:SignalPwrDef}
P_{\mathsf{signal}}(\mathbf{c}) &=& \beta_{11} \mathbf{c}^H \mathbf{h}_{11} \mathbf{h}^H_{11} \mathbf{c} \\
\label{eq:NoisePwrDef}
P_{\mathsf{noise}}(\mathbf{c})  &=& \sigma^2 \mathbf{c}^H \mathbf{c} \\
\label{eq:PilotInterPwrDef}
P_{\mathsf{contam}}(\mathbf{c}) &=& \mathbf{c}^H \left( \sum^B_{j=2} \beta_{j1} \mathbf{h}_{j1}\mathbf{h}^H_{j1} \right) \mathbf{c} \\
\label{eq:InterPwrDef}
P_{\mathsf{inter}}(\mathbf{c})   &=& \mathbf{c}^H \left( \sum^B_{j=1} \sum^K_{k=2}\beta_{jk}\mathbf{h}_{jk}\mathbf{h}^H_{jk} \right)\mathbf{c} 
\end{eqnarray}
The received SINR is then given by the expression,
\begin{eqnarray}
\label{eq:expressionSINR}
\mathsf{SINR} = \frac{P_{\mathsf{signal}} (\mathbf{c})}{P_{\mathsf{noise}}(\mathbf{c}) + P_{\mathsf{contam}}(\mathbf{c}) + P_{\mathsf{inter}}(\mathbf{c})}
\end{eqnarray}

The motivation to define $P_{\mathsf{contam}}(\mathbf{c})$ as in equation~(\ref{eq:PilotInterPwrDef}) is because the user $1$ of other cells is sending pilots in the same resource element as the user $1$ of the first base station. Hence user $1$ for $j = 2, \dots B$ contribute to the interference in a different way as compared to the rest of the users in the system. This interference contribution is termed as pilot interference and is due to pilot contaminated channel estimate being used to design linear filters. As per the definition the $P_{\mathsf{contam}}(\mathbf{c})$ is dependent on the linear filter $\mathbf{c}$ and could be different for $\hat{\mathbf{h}}_{11}$ and $\filmmsePil$.

Define $\Beta_j$ as the random variable representing the large scale fading gain from an arbitrary user in the $j$th cell. Therefore, $\beta_{jk}$ can be interpreted as the realization of $\Beta_j$ for the user $k$ and let $\Beta = \sum^B_{i=1} \Beta_i$. Next, we state the main theorem of the paper which gives the expression of SINR for a large system when an MMSE filter with a pilot contaminated estimate is used to decode the received signal.
\begin{theorem} 
\label{th:SINRMmseLargeSystem}
As $M,K \rightarrow \infty$, with $K/M = \alpha$, the SINR at the output of filter $\filmmsePil$ given in equation~(\ref{eq:MmseFilterPilot}) converges almost surely to
\begin{eqnarray}
\label{eq:SINRMmseLargeSystem}
\sinrmmsePil = \frac{ \frac{\beta_{11}}{1 +  \left(\sum^B_{j=2} \beta_{j1} \right)/\beta_{11}}}{ \sigma^2  + \frac{\left( \sum^B_{j=2} \beta^2_{j1}\right)/\beta_{11}}{ 1 + \left(\sum^B_{j=2} \beta_{j1}\right)/ \beta_{11}} + \alpha (\mathbb{E}[\Beta] -\mseInt )} 
\end{eqnarray}
where, the constants $\mseInt$, $\eta_1$, $\eta_2$ are given by
\begin{eqnarray}
\label{eq:InterSuppressionTerm}
\mseInt &=&   \mathbb{E} \left[ \frac{ \left( \frac{\Beta^2_1}{\Beta} \right)^2 \eta_1}{1 + \frac{\Beta^2_1}{\Beta} \eta_1}\right] + \frac{\eta_2}{\eta_1} \mathbb{E}  \left[  \frac{  \frac{\Beta^2_1}{\Beta} \left(\sum^B_{j=2} \frac{\Beta^2_j}{\Beta} \right) }{1 + \frac{\Beta^2_1}{\Beta}\eta_1} \right]  \nonumber \\
&& \quad  \quad  + \frac{\eta_2}{\eta_1}\mathbb{E}  \left[  \frac{  \frac{\Beta^2_1}{\Beta} \left(\sum^B_{j=2} \frac{\Beta^2_j}{\Beta} \right)}{\left( 1 + \frac{\Beta^2_1}{\Beta} \eta_1\right)^2} \right], \\
\eta_1  &=& \left( \sigma^2 + \alpha \mathbb{E}[\Beta] - \alpha \mathbb{E} \left[ \frac{ \left( \frac{\Beta^2_1}{\Beta} \right)^2 \eta_1}{1 + \frac{\beta^2_1}{\beta} \eta_1}\right] \right)^{-1}, \\
\eta_2  &=&  \left(\eta^{-2}_1 -  \alpha \mathbb{E} \left[ \left(\frac{\frac{\Beta^2_1}{\Beta}}{1 + \frac{\Beta^2_1}{\Beta} \eta_1} \right)^2\right] \right)^{-1}.
\end{eqnarray}
\end{theorem} 
\begin{IEEEproof}
Proof given in Appendix \ref{Appendix:ProofSINRMMSELargeSys}.
\end{IEEEproof}
We will see in a large system that Theorem~\ref{th:SINRMmseLargeSystem} characterizes the effect of pilot interference power and  interference averaging. Specifically, in order to put Theorem~\ref{th:SINRMmseLargeSystem} into proper perspective we state two propositions which are the results for SINR  of a large system with MMSE filter employing a perfect estimate and a matched filter with pilot contaminated estimate respectively. 
\begin{proposition} 
\label{th:SINRMmsePerLargeSystem}
As $M,K \rightarrow \infty$, with $K/M = \alpha$, the SINR at the output of filter $\filmmsePer$ given in equation~(\ref{eq:MmseFilterPer}) converges almost surely to
\begin{eqnarray}
\label{eq:SINRMMsePerLargeSystem}
\sinrmmsePer = \beta_{11} \eta_1  = \frac{\beta_{11}}{ \sigma^2 + \alpha \sum^N_{j=2}\mathbb{E}[\Beta_j] + \alpha\mathbb{E} \left[ \frac{\Beta_1}{1 + \Beta_1 \eta^*_1}\right]}
\end{eqnarray}
where, $\eta^*_1 =  \left( \sigma^2 + \alpha \sum^B_{j=2} \mathbb{E}[\Beta_j] + \alpha \mathbb{E}\left[ \frac{\Beta_1}{1 + \Beta_1 \eta^*_1}\right] \right)^{-1}$.
\end{proposition} 
\begin{IEEEproof}
We state the proposition without proof as it is straightforward to obtain it from the large system analysis techniques used for CDMA systems in \cite{TseHanly1999,JEvans:IT:2000}.
\end{IEEEproof}
$\sinrmmsePer$  is the SINR with MMSE filtering with a perfect channel estimate to its own users. This is best case scenario as compared to the MMSE with a channel estimate. We do not expect the SINR of MMSE filter with estimate to exceed this $\sinrmmsePer$. 

\begin{proposition} 
\label{th:SINRLargeSystem}
As $M,K \rightarrow \infty$, with $K/M = \alpha$, the SINR at the output of filter $\mathbf{c} = \mathbf{\hat{h}}_{11}$ converges almost surely to
\begin{eqnarray}
\label{eq:SINRLargeSystem}
\sinrmatPil = \frac{    \frac{ \beta_{11} }{ 1 +  \left( \sum^{B}_{j=2}\beta_{j1} \right)/\beta_{11} }   }{ \sigma^2 +     \left[  \frac{ \left(\sum^{B}_{j=2}\beta^2_{j1} \right)/\beta_{11}}{1 +  \left(\sum^{B}_{j=2} \beta_{j1} \right)/\beta_{11} } + \alpha \mathbb{E}[\Beta]\right]  }.
\end{eqnarray}
\end{proposition} 
\begin{IEEEproof}
Proof given in Appendix \ref{Appendix:ProofSINRMatchedLargeSys}.
\end{IEEEproof}
It is interesting to see that the expression for $\sinrmatPil$ converges to a similar expression to the result in matched filtering  $\sinrmmsePil$. If $\SignalPwr = \beta^2_{11}/ \sum^B_{j=1} \beta_{j1}$, $\PilotPwr =\sum^{B}_{j=2}\beta^2_{j1}/\sum^B_{j=1} \beta_{j1}$ and 
\begin{eqnarray}
  \InterPwr(\mathbf{c})= 
\begin{cases}
    \mathbb{E}[\Beta],& \mathbf{c} = \hat{\mathbf{h}}_{11}\\
    \mathbb{E}[\Beta] - \mseInt, & \mathbf{c} = \filmmsePil
\end{cases}
\end{eqnarray}
then we can define the expression for the asymptotic SINR with matched filtering and MMSE filtering with pilot contaminated channel estimate as,
\begin{eqnarray}
\mathsf{SINR}(\mathbf{c}) = \frac{\SignalPwr}{ \sigma^2 + \PilotPwr + \alpha \InterPwr (\mathbf{c})}.
\label{eq:SINRGeneralizedExp}
\end{eqnarray}
Here, $\SignalPwr$ is the effective signal power, $\PilotPwr$ is termed as the pilot interference power and is the consequence of employing pilot contaminated channel estimate. In the same lines as in \cite{TseHanly1999}, the filter dependent term $\InterPwr(\mathbf{c})$ is called the interference averaging term and is relevant when $\alpha \neq 0$. Although, $M \rightarrow \infty$ implying that the channel between the user $1$ is asymptotically orthogonal to channel between any other users in the system, because $K \rightarrow \infty$ with $K/M = \alpha$, the contribution to sum interference from all the users is non-zero and is given by $\alpha \InterPwr(\mathbf{c})$ for a linear filter $\mathbf{c}$. If $\nu_1 = \left( \frac{\beta_{11}}{\beta^{(1)}} \right)^2$, then the following equations shows the relationships between the $\SignalPwr$ and $\PilotPwr$ to that of definitions in equation~(\ref{eq:SignalPwrDef}) and (\ref{eq:PilotInterPwrDef}):
\begin{eqnarray}
  \frac{P_{\mathsf{signal}}(\hat{\mathbf{h}}_{11})}{\sqrt{\nu_1}} &=& \frac{P_{\mathsf{signal}}(\filmmsePil) }{\SignalPwr \eta^2_1  } = \SignalPwr \\
 \frac{P_{\mathsf{contam}}(\hat{\mathbf{h}}_{11})}{\sqrt{\nu_1}} &=& \frac{P_{\mathsf{contam}}(\filmmsePil)}{\SignalPwr \eta^2_1} = \PilotPwr
\end{eqnarray}
It is seen that for both the filters $\hat{\mathbf{h}}_{11}$ and $\filmmsePil$, their respective signal powers given by $P_{\mathsf{signal}}(\hat{\mathbf{h}}_{11})$ and $P_{\mathsf{signal}}(\filmmsePil)$ are just scaled versions of $\SignalPwr$. Similarly, $P_{\mathsf{contam}}(\hat{\mathbf{h}}_{11})$ and $P_{\mathsf{contam}}(\filmmsePil)$ are the scaled versions of $\PilotPwr$. The following can be concluded for both matched filter as well as MMSE filter with a pilot contaminated estimate:
\begin{itemize}
\item The contribution of the pilot contaminated channel estimate to interference is given by $\PilotPwr$ is same for both matched filter and MMSE filter. 
\item The contribution of pilot interference is independent of $\alpha$ and is equal to $\PilotPwr$ for all values of $\alpha$. 
\end{itemize}
As trivial as the above comments may seem it is not obvious for the filter $\filmmsePil$ from the definition of filter dependent pilot interference power $P_{\mathsf{contam}}$ in equation~(\ref{eq:PilotInterPwrDef}). This is because the matrix $\mathbf{S}$ in filter $\filmmsePil$ is dependent upon the channel of all the users in the system through the pilot contaminated channel estimate $\hat{\mathbf{h}}_{1k}$ for all $k = 1, \dots, K$. However as we see in the appendix \ref{Appendix:ProofSINRMMSELargeSys}, the contribution of the matrix $\mathbf{S}$ can be summed up into the constant $\eta_1$ for a large system. Further, the following relations can also be obtained on the interference averaging term $\InterPwr(\mathbf{c})$ when $\mathbf{c} = \hat{\mathbf{h}}_{11}$ and $\mathbf{c} = \filmmsePil$:
\begin{eqnarray}
 \frac{P_{\mathsf{noise}}(\filmmsePil) +  P_{\mathsf{inter}}(\filmmsePil)}{\SignalPwr \eta^2_1 } &=& \sigma^2 + \alpha \InterPwr(\filmmsePil), \\
 \frac{P_{\mathsf{noise}}(\hat{\mathbf{h}}_{11}) +  P_{\mathsf{inter}}(\hat{\mathbf{h}}_{11})}{\sqrt{\nu_1}}  &=& \sigma^2 + \alpha \InterPwr(\hat{\mathbf{h}}_{11}).
\end{eqnarray}

We have shown that $\InterPwr(\hat{\mathbf{h}}_{11})  - \InterPwr(\filmmsePil) = \mseInt \geq 0$  implying that interference suppression of amount $\mseInt$ can be always achieved. The amount of interference suppression $\mseInt$ is of course depended on $\alpha$ through equation~(\ref{eq:InterSuppressionTerm}).

The SINR expression in the limit of infinite antennas but finite number of users per cell are obtained when we put $\alpha = 0$ in equations~(\ref{eq:SINRMmseLargeSystem}) and (\ref{eq:SINRLargeSystem}) or in (\ref{eq:SINRGeneralizedExp}). This corresponds to a similar expression as for downlink SINR in \cite[Equation 16]{Jubin}. Since the SNR of the uplink received signal in equation~(\ref{eq:RcvdSgnlVectorForm}) is given by $\mathsf{SNR} = 1/\sigma^2$, it is seen that the SINR expression so obtained is limited by the pilot interference powers at high SNRs and we obtain \cite[Equation 13]{Marzetta2011}. This need not necessarily be achieved by transmitting at high power per symbol or equivalently by having $\sigma^2 \rightarrow 0$. Had the transmission power of data symbols be not scaled by the number of antennas, the SNR of the received signal would be linearly increasing in $M$ and when $\alpha = 0$ we again obtain \cite[Equation 13]{Marzetta2011}.

The deterministic equivalent for SINR in Rayleigh fading for MMSE filter and the corresponding expression $\mseInt$ representing the interference suppression power are our main contribution of this paper. The terms $\mathbb{E}[\Beta]$ and $\mseInt$ can be computed offline for a system with the knowledge of large scale fading distribution. It can also be estimated without the knowledge of the large scale fading distribution from user realizations over time. Depending on the value of $\mathbb{E}[\Beta] - \mseInt$ and the operating point $\alpha$ a decision can be made whether to use an MMSE filter or an matched filter. As we will see in the next section, the interference suppression obtained with an MMSE filter is necessary in increasing the outage SINR and achievable rate of the system, when there are considerable number of users as represented by the ratio $\alpha > 0.1$. This is as opposed to the regime in which antennas far outnumber users. In this operating point $\alpha \approx 0$, and then MMSE filter itself may not be necessary as pilot signals are the main contributor to interference. This regime with $\alpha \approx 0$ has been well explored in recent studies in \cite{Marzetta2011,Marzetta:2013,Hoydis:2013,Ngo:2011}. As mentioned earlier most of our focus for performance analysis is on the regime $\alpha> 0.1$ although the results are perfectly valid for any $\alpha \geq 0$. Across the users in the system the SINR is a random variable by virtue of different received powers of both the signal and the interferers contributing to pilot contamination. Also, the pilot interference power is random by virtue of the choice of the interferers contributing to pilot contamination. 

\section{Performance Analysis}
\begin{figure}[t]
    \begin{center}
      \includegraphics[width = 3.5in, clip = true, trim = 10 0 10 0]{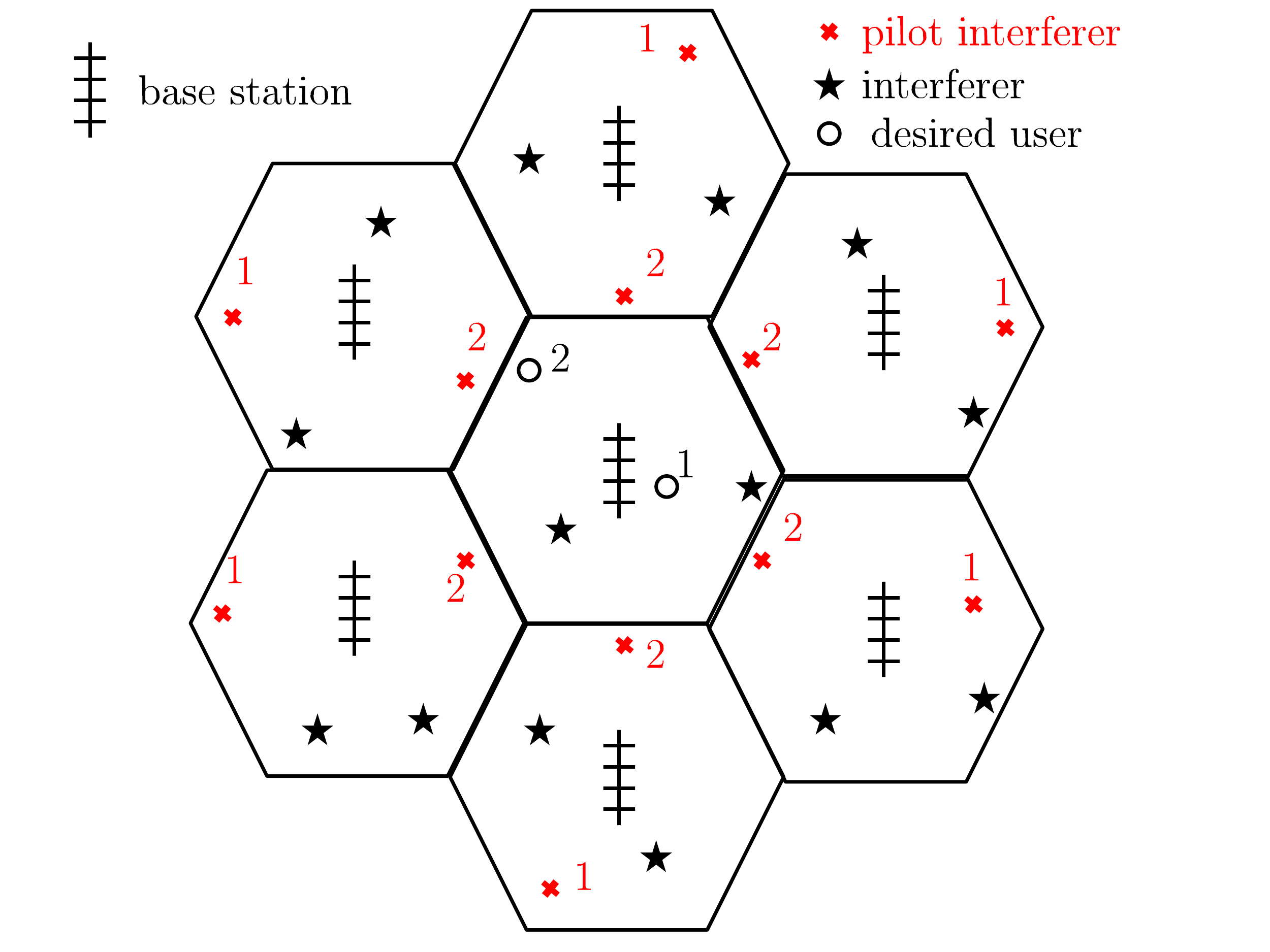}
      \caption{ In the favourable case, the sum of the received powers of interferers contributing to pilot contamination are very less as compared to that of the desired user. User $1$ in the center cell represents such a scenario. The SINR with a pilot corrupted estimate is then comparable to that of perfect estimate. On the other hand for user $2$ in the center cell, the pilot interferers received powers are comparable to that of the desired user and represents the worst case scenarios.} 
      \label{fig:7cellSys}
    \end{center}
\end{figure}
\begin{figure}[t]
    \begin{center}
      \includegraphics[width = 3.5in, clip = true, trim = 100 240 90 240]{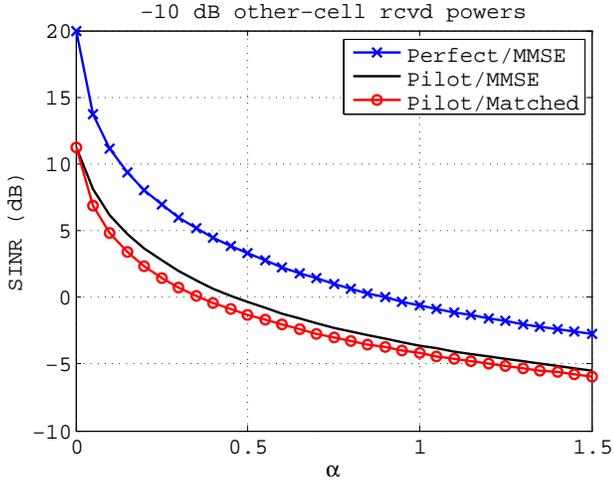}
      \caption{ The plot shows the asymptotic SINR of the first user in the first cell when MMSE filter with pilot contaminated estimate is used to decode the received signal along with the baseline comparison criterion of MMSE with a perfect estimate and matched filter with a pilot contaminated estimate for an idealized seven cell set up. It is seen that when the other cell received power is just $10$~dB lower than that in cell users the MMSE filter with pilot contaminated estimate performs close to its corresponding matched filter. This is because the limitation is now the other cell interference which MMSE filter is not designed to suppress. Hence, we  do not expect the MMSE filter with pilot estimate to be useful when $\beta_j >0.1$ } 
      \label{fig:m50a1}
    \end{center}
\end{figure}
\begin{figure}[t]
    \begin{center}
      \includegraphics[width = 3.5in, clip = true, trim = 100 240 100 240]{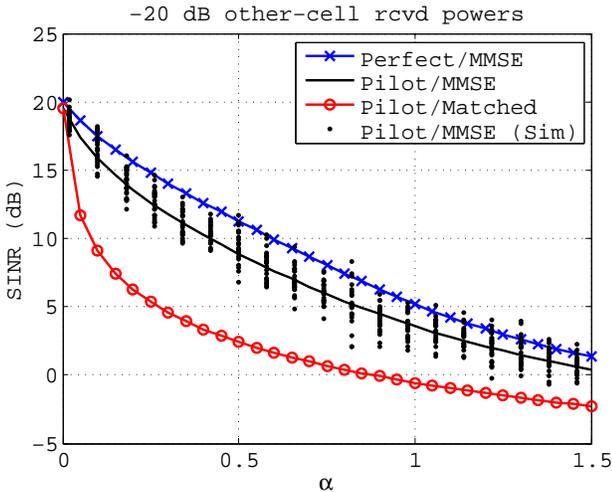}
      \caption{The plot shows the asymptotic SINR of the first user in the first cell when MMSE filter with pilot contaminated estimate is used to decode along with the baseline comparison criterion of MMSE with a perfect estimate and matched filter with a pilot contaminated estimate for an idealized seven cell set up. In this case when the other cell received powers is $20$~dB lower than that of in-cell received powers significant gains are obtained as compared to a matched filter with pilot contaminated estimate } 
      \label{fig:m50a01}
    \end{center}
\end{figure}

For the numerical evaluation, we consider hexagonal cells with users uniformly distributed in each of the cells, as shown in Fig. \ref{fig:7cellSys}. We consider a scenario where $6$ closest cells are interfering with the center cell. We assume $\beta_{1k} = 1$ so that received powers from all the users within a cell are unity.  We consider a high SNR of $20$~dB and the received powers from all the users in other cells are assumed to take a constant value of $\beta_{jk}= 0.001$, or  $0.01$, or $0.1$ for $j \neq 1$. These represent the contribution of other cell interference for three different idealized scenarios.  The interference from other cells is strong as $\beta_{jk}$ is close to $1$. We consider the SINR for the user one in the center cell. Figs.~\ref{fig:m50a1},~\ref{fig:m50a01} plot the asymptotic SINR of the MMSE with a pilot corrupted estimate given by $\sinrmmsePil$ for the case of different received powers. Although in theory the effect for small scale fading vanishes only with infinite number of antennas it is seen in Fig.~\ref{fig:m50a01} that even for a $50$-antenna base station the actual SINR realizations obtained through simulations are very near to the  asymptotic limit. We also plot $\sinrmatPil$ and $\sinrmmsePer$ as baseline for performance comparisons. In Fig.~\ref{fig:m50a1}, it is seen that $\sinrmmsePer$ is already affected by other cell interference due $\beta_{jk} = 0.1$ for $j \neq 1$. Hence, the $\sinrmmsePil$ is not expected to perform better than that and there is further $4$~dB loss due to pilot contamination. However, in the other extreme case when the other cell $\beta_{jk}$'s are close to zero, the channel estimate is already better and the $\sinrmmsePil$ performs close to $\sinrmmsePer$. Useful gains employing an MMSE filter with pilot contaminated channel estimate can be obtained when the other cell $\beta_{jk}$'s are neither close to zero or close to unity. In this example when the $\beta_{jk}= 0.01$ for $j \neq 1 $, around $7$~dB gains are possible  in comparison with matched filter with pilot estimate when operating at $\alpha = 0.5$ as seen from fig.~\ref{fig:m50a01}. While there is a loss of $3$~dB with respect to the perfect MMSE due nature of channel estimate, the reader is reminded that this is a worse case loss. The curves closes in as we decrease $\alpha$ which represents the $M \gg K$ scenario and also when $\alpha$ increases as in that case the limitation is now the averaged interference term.

\begin{figure}[t!]
    \begin{center}
      \includegraphics[width = 3.5in, clip = true, trim = 100 240 100 240]{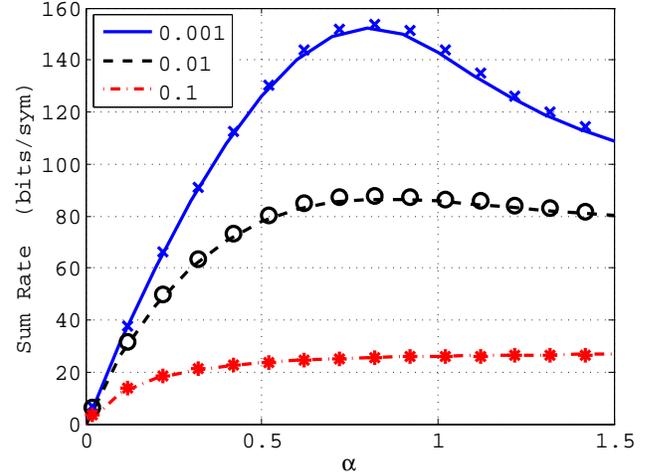}
     \caption{The plot shows the sum rate of the users in first cell for different operating points of $\alpha$ corresponding to a SNR of $20$~dB. The three curves correspond to three different received powers from other cell users. The markers correspond to simulation with $50$-antenna base stations and match the theoretical predictions.} 
      \label{fig:AchRate}
    \end{center}
\end{figure}

In fig.~\ref{fig:AchRate}  we plot the achievable sum rate for users in the first cell with each users SNR being $20$~dB. We assume large enough coherence time so that the training time need not be taken into account. This is because our focus is on the sum rate achievable with variation in $\alpha$. However, if necessary the sum rate can be easily adjusted based on training overhead when coherence time is a significant factor. We fix the number of antennas and calculate the sum rate with varying $\alpha$ as $\alpha M \log_2(1 + \sinrmmsePil)$. The three curves corresponds to the received powers of all users from other cells  being either $\beta_{jk}= 0.001$, or $0.01$, or $0.1$ for j $\neq 1$ assuming unit received power from the in-cell users. Sum rates of over $20$~bits/symbol are achieved for users when the other cell received powers are below $10$~dB of the in-cell received powers. Also, the simulation with $50$-antenna base station is seen to match the theoretical rates predicted for this set up. The interference limited system has the flexibility to serve a large number of users at low SINR or a few number of users at a high SINR depending on the operating point $\alpha$.  The plot suggest a optimal operating point $\alpha$ for which the sum rate is maximum. For example when $\beta_j =0.01$ and $\alpha = 0.8$ gives a sum rate of around $88$~bits/symbol. Larger $\alpha$ causes the $\sinrmmsePil$ to be lower so that the $\alpha$ term outside the $\log_2$ is ineffective to increase sum rate while a lower $\alpha$ implies that less users are served and hence lesser sum rate. When the other-cell received powers are large, the curve flattens and the sum rate is constant for most of the operating points $\alpha$.

\begin{figure}[t]
    \begin{center}
      \includegraphics[width = 3.5in, clip = true, trim = 100 240 100 240]{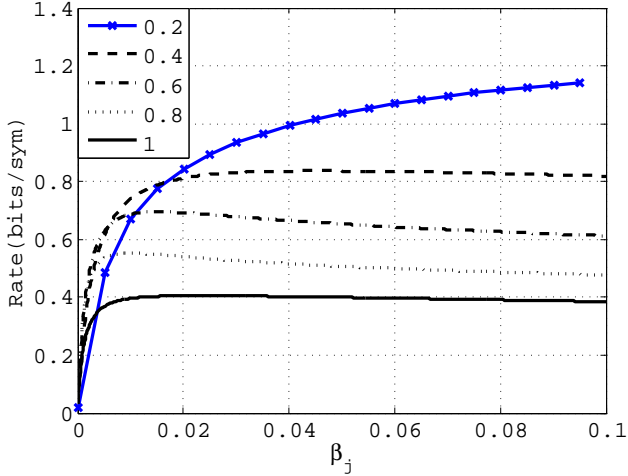}
      \caption{ The plot shows the loss of rate due to pilot contamination  when an MMSE filter with pilot contaminated estimate is used to decode the received signal as compared to perfect estimate MMSE filter. The different curves correspond to different values of $\alpha$ from $0.2$ to $1$. It is seen that smaller the $\alpha$ the MMSE filter with pilot contaminated estimate performs worse than the ideal MMSE filter. However, the sum rate for users per base station is different.} 
      \label{fig:RateDiff}
    \end{center}
\end{figure}

In fig.~\ref{fig:RateDiff},  we plot the difference of achievable rate per user between MMSE filter with a perfect estimate and MMSE filter with a pilot estimate for different values of other cell interference power. We do not take into account the training overhead for comparison assuming we obtain the perfect estimate with the same training time. We limit ourselves to $\beta_j < 0.1 $ since $\sinrmmsePil$ is already close to $\sinrmatPil$ otherwise. Also in $\beta_j > 0.1$ regime, there is significant other cell interference which both the perfect estimate based and the pilot based MMSE filter are not designed to suppress, thereby  affecting the achievable rates. We plot five different curves corresponding to system operating points $\alpha$.  As seen earlier the total interference with filter $\filmmsePil$ is given by $\PilotPwr + \alpha \InterPwr(\filmmsePil)$ and with $\filmmsePer$ the interference power is given by $\alpha \left( \sum^B_{j=2}\mathbb{E}[\Beta_j] + \mathbb{E}\left[ \frac{\Beta_1}{1 + \Beta_1 \eta^{*}_1}\right] \right)$. When $\alpha$ is small then the significant loss of rate with $\filmmsePil$ is due to $\PilotPwr$ as the filter dependent interference power for both filters in negligible. On the hand, when $\alpha$ is large $\alpha \InterPwr$ dominates the $\PilotPwr$ and since both the filter are affected by $P_{\mathsf{inter}}(\mathbf{c})$ the rate difference is small. Although when $\alpha = 1$ there is only a $0.4$~bits/symbol difference the sum rate will be affected differently. For example in a $50$-antenna base station with $50$ users at $\beta_j = 0.1$ this could mean that sum rate with pilot contaminated MMSE filter is $20$~bits/symbol  lower than that perfect MMSE filter. On the other hand when $\alpha = 0.2$, the sum rate difference is $12$~bits/symbol.

\subsection{Effect of Pilot Contamination}
Through a couple of typical possible realizations of user positions, we explain the effect of pilot interference in $\sinrmmsePil$, $\sinrmatPil$ and compare it with that of the SINR with a perfect estimate. For illustration, in Fig.~\ref{fig:7cellSys} consider only distance based pathloss in large scale fading although the result holds when shadowing is also present. This is applicable to both matched filter and MMSE filter. Consider the first scenario when
\begin{eqnarray}
\label{eq:favScen}
\frac{\sum^{B}_{j=2} \beta_{j1}}{\beta_{11}} \ll 1 \Rightarrow  \frac{\sum^{B}_{j=2} \beta^2_{j1}}{\beta_{11}} \ll1
\end{eqnarray}
This corresponds to the fact that sum of received powers of the interferers are much less that that of desired user power. Under these conditions the SINR of the received signal in equation (\ref{eq:SINRLargeSystem}) is,
\begin{eqnarray}
\mathsf{SINR} &\approx &    \frac{ \beta_{11} }{ \sigma^2 +      \alpha (\mathbb{E}[\Beta] - \mathcal{I})} \nonumber
\end{eqnarray}
where, $\mathcal{I} = 0$ if matched filter is employed or $\mathcal{I} = \mseInt $  if MMSE filter $\filmmsePil$ is employed. As we will see in the next section, typically scenarios show that the interference suppression power $\mseInt$ is almost same as what could have been with a filter $\filmmsePer$. Hence the SINR of the filter with the corrupt channel estimate is as good as the SINR with a perfect channel estimate.  In Fig.~\ref{fig:7cellSys}, the situation of user $1$ in the center cell represents the favourable scenario with the interferers contributing to the pilot contaminated channel estimate are far such that the condition (\ref{eq:favScen}) is satisfied. On the other hand if gains of all the interferers are comparable to that of the desired users, i.e.,
\begin{eqnarray}
\frac{\sum^{B}_{j=2} \beta_{j1}}{\beta_{11}} \approx B - 1
\end{eqnarray}
then pilot interference contributes negatively to the SINR in addition to interference averaging. This is represented by realization of user $2$ of the center cell in Fig.~\ref{fig:7cellSys}. Therefore, we can conclude that, as compared to the linear filter with perfect estimate, the filter with a pilot estimate has higher probability that it is less than a given SINR. 

\subsection{Five Percentile SINR}
\label{subsec:FivePer}
\begin{figure}[t!]
    \begin{center}
      \includegraphics[width = 3.5in, clip = true, trim = 100 240 100 240]{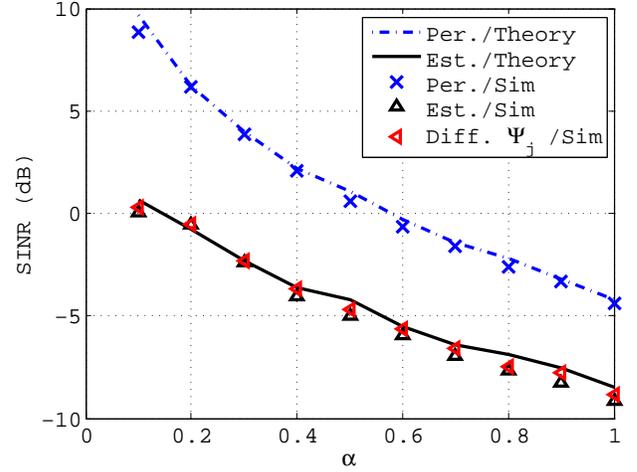}
      \caption{ The plot shows the five percentile SINR of the first user in the first cell for a seven cell set up in a non-idealized scenario. The received powers from users can be different depending on their positions and shadowing and hence the received SINR is random. The theoretical curves are matched with simulation. The details are described in section \ref{subsec:FivePer}. } 
      \label{fig:VaryAlpha}
    \end{center}
\end{figure}
In the earlier section we showed that pilot contamination has the effect of reducing the outage SINR. In order to get more intuition under practical scenarios of large scale fading gains, we consider the seven cell model with cell radius is $R = 1$~km, and assume a COST231 model for propagation loss between the base station and the users. The noise power is assumed to be $-174$~dBm and user transmit power of $23$~dBm. We plot the five percentile of the SINR in Fig. \ref{fig:VaryAlpha} for the perfect MMSE filtering given by $\sinrmmsePer$ and MMSE filter with pilot contamination given by $\sinrmmsePil$, for varying values of $\alpha$. To that extend, we compute the interference terms $\mathbb{E}[\Beta]$ and $\mseInt$ offline by averaging over a sufficient number of user positions. Also, $\eta_1$ and $\eta_2$ can be computed offline for different values of $\alpha$ as they are constant for the system and dependent on the large scale fading characteristics. Notice that $\sinrmmsePil$ is devoid of the small scale fading parameters. Also, for $\sinrmmsePer$ we compute the terms, $\eta^*_1$ and the average interference $\mathbb{E}[\Beta] - \mathbb{E}\left[ \frac{\Beta^2_1 \eta^*_1}{1 + \Beta_1 \eta^*_1}\right]$. It is found that $\mathbb{E}[\Beta] - \mathcal{C} = 38$~dB and $\mathbb{E}[\Beta] - \mathbb{E}\left[ \frac{\Beta^2_1 \eta^*_1}{1 + \Beta_1 \eta^*_1}\right] = 36$~dB. This shows that in terms of the interference suppression the performance of both filters $\filmmsePer$ and $\filmmsePil$ are almost same.  In order to compare the theoretical expression we also plot the five percentile SINR which is generated using simulations. These involves computing the SINRs for various small scale fading channel realizations along with large scale fading. For the simulation we use $50$ antenna base stations each serving a different number of users corresponding to different $\alpha$. The channel estimate is based on same in-cell orthogonal training sequences being repeated across the cells. Further, we also compare the five percentile SINR obtained out of actual channel estimation. In order for that we assume different independently generated in-cell orthogonal training sequences which are non-orthogonal across the cells. We perform an MMSE estimation of the channel and generate MMSE filter  using the actual channel estimate given by equation~(\ref{eq:trainingBasedEst}).

It is seen through Fig.~\ref{fig:VaryAlpha}, that $\sinrmmsePil$ in equation~(\ref{eq:SINRMmseLargeSystem}) matches the five percentile SINR obtained through simulation and is typically less by about $0.3$~dB of the theoretical expression. This is true for both in-cell pilot sequences being repeated across the cells as well as different and independent  pilot training sequences across the cells. This also implies that the even a $50$ antenna base station is large enough for the theoretical predictions to be effective in addition to being independent of the effect of small scale fading in the resulting SINR. As we increase the number of antennas the theoretical expression exactly matches the SINR obtained through simulation. Also, the MMSE filter with pilot contamination performs just $5$~dB below the MMSE filter with perfect channel estimate and this gap is unambiguously a result of the pilot contaminated channel estimate. Also it is seen that even at $\alpha = 1$, which implies a heavily loaded system the five percentile $\sinrmmsePil$ is $-9$~dB which is well within the sensitivity of base station receivers.
\begin{table}[t!]
\begin{center}
\begin{tabular}[c]{ | c | c | c | c | c | c |}
\hline
\multirow{2}{*}{$\alpha$}        & \multicolumn{2}{| c |}{$M = 50$} &  \multicolumn{2}{| c |}{$M = 10$}\\ \cline{2-5}              
   &  $R_{\mathsf{per}}$   & $R_{\mathsf{pilot}}$ &  $R_{\mathsf{per}}$   & $R_{\mathsf{pilot}}$ \\ \hline
 $0.1$  & $6.0$ & $4.7$ &   &  \\
 $0.2$  &  $5.0$ &  $4.0$ &  &  \\
 $0.3$  &  $4.3$&  $3.4$ & $4.4$ & $3.6$ \\
 $0.4$  & $3.8$ & $3.0$ & $4.0$& $3.2$\\
 $0.5$  & $3.4$& $2.7$ & $3.6$ & $2.9$\\ 
 $0.6$  & $3.1$ & $2.5$ & $3.3$ & $2.7$\\
 $0.7$  & $2.8$ & $2.3$ & $3.0$ & $2.4$\\
 $0.8$  & $2.64$ & $2.1$ & $2.7$ & $2.2$\\
 $0.9$  & $2.4$ & $1.9$ & $2.4$ & $2.0$\\
 $1.0$  & $2.2$ & $1.9$ & $2.3$ & $2.0$\\ \hline
\end{tabular}
\end{center}
\caption{The table shows the achievable rate (bits/symbol)  with MMSE filtering for a $50$ and $10$-antenna base station serving different number of users. $R_{\mathsf{per}}$ corresponds to rate with a perfect estimate and $R_{\mathsf{pilot}}$ corresponds to the pilot contaminated filter.}
\label{tab:AchRate}
\end{table}

Table~\ref{tab:AchRate} shows the achievable rates per symbol for a user in the central cell using MMSE based detection with perfect estimate and pilot contaminated estimate. The achievable rate is given by $R = \mathbb{E}[\log (1 + \sinrmmsePil)]$ which here is calculated by averaging the instantaneous rate over $2 \times 10^3$ realizations of user positions for user $1$. This is the same for all users in the central cell. Both theory and simulations based on $50$ base stations antennas serving different number of users agree to the numbers shown in the table. It is seen that the difference between them is approximately $1$~bit/symbol for small $\alpha > 0.1$ and closes in when it increases. However, for $\alpha \ll 1 $, the difference between the rates increases as effect of pilot interference will never let the SINR approach SNR.

Further, Table~\ref{tab:AchRate}, also shows the simulated results for the achievable rate for a $10$-antenna base station with $3$ to $10$ users. It is seen that even for a $10$ antenna base station, the simulated results agree closely to the earlier results obtained from theory in Table~\ref{tab:AchRate}. This highlights the usefulness of the large system analysis in providing accurate predictions for achievable rates for not necessarily large number of antennas but also contemporary MIMO systems. However, we would like to point out that with more number of antennas we can serve more users at the same rate given below.

\section{Conclusion}
\label{sec:conclusion}
In this work we found the expression for SINR for a large system when a MMSE filter with a pilot contaminated estimate is employed to decode the received signal. We validated the expression through simulations and showed that a $50$-antenna base station serving different number of users is sufficient enough to employ our large system results. We characterized the effect of pilot contamination in that it has the effect of reducing the five percentile SINRs as compared to the MMSE with perfect estimate for all values of $\alpha$. We also found an explicit expression for the interference suppression power due to MMSE filter and compared it with that of matched filter. We showed that five percentile SINR of the MMSE with a pilot contaminated estimate is within $5$~dB of MMSE filter with a perfect estimate. We also found that the results with actual channel estimation match the theoretical results. In this work we have assumed that the training time allocated to users is $K$ symbols. In future work, we wish to study the training overhead for different values of training time and users depending on the coherence time. We believe this can be done through a large system analysis of training time versus the number of users per cell. It would also be interesting to see if the considerable work done by the authors in \cite{Hoydis:2013}, in getting a generalized expression for the deterministic equivalent of the SINR can be further simplified into intuitive expressions for other channel models. This will be of help in realizing engineering conclusions tailored for different channels models like distributed antennas, correlated antennas, distributed sets of correlated antennas \cite{JZHANG:2013} to name a few.  Also, recent work has proposed that pilot contamination as an artefact of linear channel estimation techniques \cite{Müller2013}. While they have provided a theoretical understanding in an ideal situation, practical solutions applicable to a regime with large number of users are still to be found. Algorithms from multi-user detection for CDMA systems are a useful tool when we have enough coherence time \cite{VerduMUD}. We are currently looking into such adaptive algorithms that could be implemented for short coherence time scenarios.

\appendices
\section{Results from Literature}
In this section we briefly describe the necessary results from literature to derive the asymptotic SINR expressions. These results were used previously in the context of CDMA systems in \cite{TseHanly1999,JEvans:IT:2000,JZhang:IT:2001}.
\begin{lemma} \cite[Lemma 1]{JEvans:IT:2000 }
\label{lemma:TraceLargeMatrix}
If $\mathbf{S}$ is a deterministic $M \times M$ matrix with uniformly bounded spectral radius for all $M$. Let $\mathbf{q} = \frac{1}{\sqrt{M}}\begin{bmatrix} q_1 & q_2 & \dots & q_M \end{bmatrix}^T$ where $q_i$'s are i.i.d complex random variables with zero mean, unit variance and finite eight moment. Let $\mathbf{r}$ be a similar vector independent of $\mathbf{q}$. Then,
\begin{eqnarray}
\mathbf{q}^H \mathbf{S} \mathbf{q}  & \rightarrow &  \frac{1}{M}  \trace \{ \mathbf{S} \} , \\
\mathbf{q}^H \mathbf{S} \mathbf{r}  & \rightarrow & 0
\end{eqnarray}
almost surely as $M \rightarrow \infty$.
\end{lemma}

Results in linear MMSE filters for large dimensions have been obtained using Stieltjes transform result on symmetric matrices in \cite{TseHanly1999, JZhang:IT:2001, JEvans:IT:2000}. For completeness and clarity of understanding of MMSE for multi-user MIMO with pilot contamination we first define the Stieltjes transform of a random variable and state the result without proof here.
\begin{definition}\cite[Section 2.2.1]{TulinoVerdu} Let a real valued random variable be given by the distribution $G$. Then, the Stieltjes transform $m(z)$ with complex argument $z$ and positive imaginary part is defined as
\begin{eqnarray}
m(z) = \int \frac{1}{\lambda - z} d G(\lambda).
\end{eqnarray}
\end{definition}
\begin{theorem}\cite[Theorem 6]{JEvans:IT:2000}
\label{th:StieltjesTransformResult}
Let $\mathbf{X} \in \mathbb{C}^{M \times K}$ be a  matrix with independent and identically distributed complex entries each with variance $1/M$. Also, let $\mathbf{T} \in \mathbb{C}^{K \times K}$ be a random hermitian non-negative definite matrix independent of $\mathbf{X}$ such that the empirical distribution of its eigenvalues converges to a fixed distribution $F$ as $M \rightarrow \infty$.  Then as $K \rightarrow \infty$ and $M \rightarrow \infty$ with $K/M = \alpha$, almost surely the empirical distribution of eigenvalues of $\mathbf{XTX}^H$ converges to a non-random distribution function $G$ whose Stieltjes transform $m(z)$ satisfies,
\begin{eqnarray}
\label{eq:StieltjesTransformResult}
m(z) = \frac{1}{-z + \alpha \int \frac{p}{1 + p m(z)} dF(p) },
\end{eqnarray}
for $z$ complex with positive imaginary part.
\end{theorem}
Next we state a corollary from \cite{JZhang:IT:2001} which is also a consequence of Stieltjes transform result in Theorem \ref{th:StieltjesTransformResult}.
\begin{corollary} Where  Theorem \ref{th:StieltjesTransformResult} is applicable,
\label{cor:JZhang}
\begin{eqnarray}
\alpha \mathbb{E}\left[ \frac{p \frac{d}{d z} m(z)}{(1 + p m(z))^2}\right] = \int \frac{\lambda}{(\lambda - z)^2}dG(\lambda)
\end{eqnarray}
\end{corollary}
\begin{IEEEproof}
\begin{eqnarray}
\alpha \mathbb{E}\left[ \frac{p \frac{d}{d z} m(z)}{(1 + p m(z))^2}\right] &\overset{(a)}{=}& -\alpha \frac{d}{d z} \nonumber \mathbb{E}\left[1 -  \frac{p m(z)}{1 + p m(z)}\right] , \nonumber \\
&=&  \frac{d}{d z}  \alpha \mathbb{E}\left[\frac{p m(z)}{1 + p m(z)}\right], \nonumber \\
&\overset{(b)}{=}&  \frac{d}{d z} (1 + z m(z)), \nonumber \\
&=&   \frac{d}{d z} \int \frac{\lambda}{\lambda - z} dG(\lambda),
\end{eqnarray}
where (a) is due to dominated convergence theorem and (b) is due to equation~(\ref{eq:StieltjesTransformResult}).
\end{IEEEproof}

\section{Proof of Theorem \ref{th:SINRMmseLargeSystem}}
\label{Appendix:ProofSINRMMSELargeSys}

\begin{figure*}[!t]
\normalsize
\setcounter{MYtempeqncnt}{\value{equation}}
\hrulefill
\setcounter{equation}{54}
\begin{eqnarray}
\trace \left \{ \mathbf{Z} \right \}   &=&  \trace \left \{ \sum^B_{j=1} \sum^K_{k=2}\beta_{jk} \mathbf{S}^{-1} \mathbf{H}_{k}\mathbf{e}_{j}\mathbf{e}^H_{j}\mathbf{H}^H_{k} \mathbf{S}^{-1}  \right \} \\
& \overset{(a)}{=}&  \sum^B_{j=1} \sum^K_{k=2} \beta_{jk} \mathbf{e}^H_{j}\mathbf{H}^H_{k} \left( \mathbf{S}^{-1} \right)^2\mathbf{H}_{k}\mathbf{e}_{j}  \\
&\overset{(b)}{=}& \sum^B_{j=1} \sum^K_{k=2} \beta_{jk} \mathbf{e}^H_j \mathbf{H}^H_k   \mathbf{S}^{-2}_k  \left( \mathbf{I} - 2\frac{ \nu_k\mathbf{H}_k \mathbf{a}_k \mathbf{a}^H_k \mathbf{H}^H_k \mathbf{S}^{-1}_k }{1 + \nu^2_k\mathbf{a}^H_k\mathbf{H}^H_k\mathbf{S}^{-1}_k\mathbf{H}_k\mathbf{a}_k}  + \right.\nonumber \\
&& \left. \frac{\nu^2_k\mathbf{H}_k \mathbf{a}_k \mathbf{a}^H_k \mathbf{H}^H_k \mathbf{S}^{-1}_k \mathbf{H}_k \mathbf{a}_k \mathbf{a}^H_k \mathbf{H}^H_k \mathbf{S}^{-1}_k }{(1 + \nu^2_k \mathbf{a}^H_k\mathbf{H}^H_k\mathbf{S}^{-1}_k\mathbf{H}_k\mathbf{a}_k)^2}\right)\mathbf{H}_k \mathbf{e}_j   \\
& = & \sum^B_{j=1} \left( \sum^K_{k=2} \beta_{jk} \mathbf{e}^H_j \mathbf{H}^H_k  \mathbf{S}^{-2}_k \mathbf{H}_k \mathbf{e}_j  - 2 \sum^K_{k=2} \beta_{jk} \frac{ \nu_k \mathbf{e}^H_j \mathbf{H}^H_k  \mathbf{S}^{-2}_k \mathbf{H}_k \mathbf{a}_k \mathbf{a}^H_k \mathbf{H}^H_k \mathbf{S}^{-1}_k \mathbf{H}_k \mathbf{e}_j}{1 + \nu_k \mathbf{a}^H_k\mathbf{H}^H_k\mathbf{S}^{-1}_k\mathbf{H}_k\mathbf{a}_k}  + \right . \nonumber \\
&&   \left. \sum^K_{k=2} \beta_{jk} \frac{\nu^2_k \mathbf{e}^H_j \mathbf{H}^H_k  \mathbf{S}^{-2}_k  \mathbf{H}_k \mathbf{a}_k \mathbf{a}^H_k \mathbf{H}^H_k \mathbf{S}^{-1}_k \mathbf{H}_k \mathbf{a}_k \mathbf{a}^H_k \mathbf{H}^H_k \mathbf{S}^{-1}_k \mathbf{H}_k \mathbf{e}_j}{(1 + \nu_k \mathbf{a}^H_k\mathbf{H}^H_k\mathbf{S}^{-1}_k\mathbf{H}_k\mathbf{a}_k)^2}  \right)  
 \label{eq:expandedTraceOfZ}
\end{eqnarray}
\setcounter{equation}{\value{MYtempeqncnt}}
\end{figure*} 

\begin{figure*}[!t]
\setcounter{MYtempeqncnt}{\value{equation}}
\hrulefill
\setcounter{equation}{61}
\begin{eqnarray}
\frac{ \trace\{ \mathbf{Z}\}}{M} & \overset{a.s.}{\longrightarrow} &  \alpha \sum^B_{j=1} \mathbb{E} \left[ \Beta_j \left(\eta_2 - 2  \frac{ \left(\frac{\Beta_1}{\Beta} \right)^2 \Beta_j \eta_1 \eta_2}{1 + \eta_1 \frac{\Beta^2_1}{\Beta} }+ \frac{ \left(\frac{\Beta_1}{\Beta} \right)^4 \Beta_j \Beta \eta_2 \eta^2_1}{ \left(1 + \eta_1 \frac{\Beta^2_1}{\Beta} \right)^2} \right)\right]  \nonumber \\
& = & \alpha \sum^B_{j=2} \mathbb{E}\left[ \Beta_j\right] \eta_2 +  \alpha \sum^B_{j=1} \mathbb{E}\left[ \Beta_j \frac{\Beta_1}{\Beta}\right] \eta_2   - \alpha \mathbb{E}\left[ \frac{ \left(\frac{\Beta_1}{\Beta} \right)^2 \sum^B_{j=1}\Beta_j \eta_1 \eta_2}{1 + \eta_1 \frac{\Beta^2_1}{\Beta} }\right]  - \alpha \mathbb{E}\left[ \frac{ \left(\frac{\Beta_1}{\Beta} \right)^2 \sum^B_{j=1}\Beta_j \eta_1 \eta_2}{\left(1 + \eta_1 \frac{\Beta^2_1}{\Beta}\right)^2 }\right] \nonumber \\
\label{eq:TraceInterMatrixLastbut1}
& = &  (\bar{\theta_1} + \bar{\theta_2})\eta_2  +  \alpha \mathbb{E}\left[ \frac{\frac{\Beta^2_1}{\Beta} \eta_2}{\left( 1 + \frac{\Beta^2_1}{\Beta} \eta_1\right)^2}\right]   - \alpha \mathbb{E}  \left[  \frac{  \frac{\Beta^2_1}{\Beta} \left(\sum^B_{j=2} \frac{\Beta^2_j}{\Beta} \right) \eta_1 \eta_2 }{1 + \frac{\Beta^2_1}{\Beta}\eta_1} \right] - \alpha \mathbb{E}  \left[  \frac{  \frac{\Beta^2_1}{\Beta} \left(\sum^B_{j=2} \frac{\Beta^2_j}{\Beta} \right) \eta_1 \eta_2 }{\left( 1 + \frac{\Beta^2_1}{\Beta} \eta_1\right)^2} \right] \\
& = &  \int^{\infty}_{0} \frac{\lambda + \bar{\theta}_1 + \bar{\theta}_2}{( \lambda + \bar{\theta}_1  + \bar{\theta}_2+ \sigma^2)^2} dG(\lambda)     - \alpha \mathbb{E}  \left[  \frac{  \frac{\Beta^2_1}{\Beta} \left(\sum^B_{j=2} \frac{\Beta^2_j}{\Beta} \right) \eta_1 \eta_2 }{1 + \frac{\Beta^2_1}{\Beta}\eta_1} \right] - \alpha \mathbb{E}  \left[  \frac{  \frac{\Beta^2_1}{\Beta} \left(\sum^B_{j=2} \frac{\Beta^2_j}{\Beta} \right) \eta_1 \eta_2 }{\left( 1 + \frac{\Beta^2_1}{\Beta} \eta_1\right)^2} \right]. 
\label{eq:TraceInterMatrix}
\end{eqnarray}
\setcounter{equation}{\value{MYtempeqncnt}}
\hrulefill
\end{figure*}

Let the overall channel matrix  representing the system be defined as,
$\mathbf{H} = \begin{bmatrix} \mathbf{H}_1 & \mathbf{H}_2  & \dots & \mathbf{H}_K \end{bmatrix}$,
where, 
$ \mathbf{H}_i = \begin{bmatrix} \mathbf{h}_{1i} & \mathbf{h}_{2i}  & \dots & \mathbf{h}_{Bi} \end{bmatrix}$.
Also define the large scale fading coefficient vector to be
$\mathbf{a}_i = \begin{bmatrix} \sqrt{\beta_{1i}} & \sqrt{\beta_{2i}} & \dots & \sqrt{\beta_{Bi}} \end{bmatrix}^T$,
and $\mathbf{e}_i \in \mathbb{C}^{B \times 1}$ as a unit vector with $1$ in the $i^{th}$ position for $i \in \{1,2, \dots, B \}$. With the above definitions the channel estimate $\hat{\mathbf{h}}_{1k} = \frac{\sqrt{\beta_{1k}}}{\beta^{(k)}} \mathbf{H}_k \mathbf{a}_k$ and $\filmmsePil = \frac{\beta_{11} } {\beta^{(1)}} \mathbf{S}^{-1} \mathbf{H}_1 \mathbf{a}_1$. Also, if $\nu_k = \left(\frac{\beta_{1k}}{{\beta^{(k)}}}\right)^2$, then the signal power, noise power, and pilot interference power, are respectively given by,
\begin{eqnarray}
\label{eq:SgnlPwrMmse}
P_{\mathsf{signal}}(\filmmsePil) &=& \beta_{11}\nu_1 \left| \mathbf{a}^H_1 \mathbf{H}^H_1 \mathbf{S}^{-1} \mathbf{H}_1 \mathbf{e}_1 \right|^2\\
\label{eq:noisePwrMmse}
P_\mathsf{noise}(\filmmsePil) &=& \nu_1 \sigma^2 \mathbf{a}^H_1 \mathbf{H}^H_1 ( \mathbf{S}^{-1})^2 \mathbf{H}_1 \mathbf{a}_1, \\
\label{eq:pilotInterPwrMmse}
P_{\mathsf{contam}}(\filmmsePil) &=& \nu_{1} \sum^B_{j=2} \beta_{j1} \left | \mathbf{a}^H_1 \mathbf{H}^H_1\mathbf{S}^{-1}\mathbf{H}_1\mathbf{e}_j  \right|^2.
\end{eqnarray}
For further analysis let us define, 
\begin{eqnarray}
\mathbf{Z} = \mathbf{S}^{-1}\left(\sum^B_{j=1}\sum^K_{k=2}\beta_{jk}\mathbf{h}_{jk}\mathbf{h}^H_{jk}\right)\mathbf{S}^{-1}
\end{eqnarray}
then the interference power is given by 
\begin{eqnarray}
\label{eq:InterPwrMmse}
P_{\mathsf{inter}}(\filmmsePil) = \nu_1 \mathbf{a}^H_1 \mathbf{H}^H_1 \mathbf{Z} \mathbf{H}_1 \mathbf{a}_1
\end{eqnarray}
Define a real block diagonal matrix $\mathbf{D}_1 \in \mathbb{R}^{(K-1)B \times (K-1)B}$ and $\mathbf{S}_1 \in \mathbb{C}^{M \times (K-1)B}$ as
\begin{eqnarray}
\mathbf{D}_1 &=& \diag \left \{ \nu_2 \mathbf{a}_2 \mathbf{a}^H_2, \nu_3\mathbf{a}_3\mathbf{a}^H_3, \dots ,  \nu_K\mathbf{a}_K\mathbf{a}^H_K \right \},  \\
\mathbf{S}_1 &=& \begin{bmatrix}  \mathbf{H}_2  & \dots & \mathbf{H}_K \end{bmatrix} \text{.}
\end{eqnarray}
Then the matrix $\mathbf{S}$ can be rewritten as,
\begin{eqnarray}
\mathbf{S} &=&  \sum^K_{k=2} \nu_k \mathbf{H}_k\mathbf{a}_k \mathbf{a}^H_k\mathbf{H}^H_k  + (\theta_1 + \theta_2 +\sigma^2) \mathbf{I}  \nonumber \\ 
&=&   \mathbf{S}_1 \mathbf{D}_1 \mathbf{S}^H_1 + (\theta_1 + \theta_2 +\sigma^2) \mathbf{I} \text{.}
\end{eqnarray}
For the matrix $\mathbf{D}_1$ there are $(K-1)(B-1)$ eigenvalues which are equal to zero and $K-1$ non-zero values given by $\frac{\beta^2_{1k}}{\beta^{(k)}}$, for all $k \in \{ 2, \dots, K\}$. Also, define $\Beta_{j}$ as the random variable representing the large scale fading gain from an arbitrary user in the $j^{th}$ cell. Therefore, $\beta_{jk}$ can be interpreted as the realization of $\Beta_j$ for the $k^{th}$ user.  Therefore, Theorem \ref{th:StieltjesTransformResult} takes the form 
\begin{eqnarray}
\label{eq:StieltjesSpecificToPaper}
m(z) = \left(-z + \alpha \mathbb{E} \left[\frac{\Beta^2_1/\Beta}{1 + \Beta^2_1 m(z)/\Beta} \right]\right)^{-1}
\end{eqnarray}
 where, the expectation is now over the joint distribution of the $\Beta_j$s and $\Beta$.

Also, notice that the spectral radius of $\mathbf{S}$ is bounded by $(\theta_1 + \theta_2 + \sigma^2)^{-1}$. Therefore, with $\Beta = \sum^B_{j=1} \Beta_j$, $\bar{\theta_1} = \alpha \sum^B_{j=2} \mathbb{E}\left[\Beta_{j}\right]$, $\bar{\theta_2} = \alpha \sum^B_{j=2} \mathbb{E}\left[\Beta_{j}\left( \frac{\Beta_{1}}{\Beta^{}}\right) \right]$ and using Lemma \ref{lemma:TraceLargeMatrix} we can conclude that,
\begin{eqnarray}
\label{eq:SignalPowerTrace}
\mathbf{H}^H_1 \mathbf{S}^{-1} \mathbf{H}_1  & \rightarrow & \frac{1}{M} \trace \{ \mathbf{S}^{-1} \} \mathbf{I}= \eta_1 \mathbf{I} ,\\
\label{eq:NoisePowerTrace}
\mathbf{H}^H_1 ( \mathbf{S}^{-1})^2 \mathbf{H}_1 & \rightarrow & \frac{1}{M} \trace \{ \mathbf{S}^{-2} \}\mathbf{I} = \eta_2 \mathbf{I}
\end{eqnarray}
almost surely as $M \rightarrow \infty$ where, if $G$ is the non-random limiting distribution of the eigenvalues $\lambda$ of the matrix $\mathbf{S}_1 \mathbf{D}_1 \mathbf{S}^H_1$, then
\begin{eqnarray}
\label{eq:eta_1}
\eta_1 &=& \int \frac{1}{\lambda + \bar{\theta_1} + \bar{\theta_2} + \sigma^2}dG(\lambda)  \text{  and }\\
\label{eq:eta_2}
\eta_2 &=& \int \frac{1}{(\lambda + \bar{\theta_1} + \bar{\theta_2} + \sigma^2)^2}dG(\lambda) \text{.}
\end{eqnarray}
From equations~(\ref{eq:eta_1}),~(\ref{eq:eta_2}) and using the definition of Stieltjes transform \cite{TulinoVerdu} we can find that $ \eta_1 = \lim_{-z \rightarrow \bar{\theta}_2 + \bar{\theta}_2 + \sigma^2} m(z) $, and since $m(z)$ is complex analytic $\eta_2 = \lim_{-z \rightarrow \bar{\theta}_2 + \bar{\theta}_2 + \sigma^2} \frac{d}{dz} m(z)$. The values of $\eta_1$ and $\eta_2$ can then also be obtained from solving equation~(\ref{eq:StieltjesSpecificToPaper}) and its derivative. The equations are given by,
\begin{eqnarray}
\eta_1  &=& \left( \sigma^2 + \alpha \mathbb{E}[\Beta] - \alpha \mathbb{E} \left[ \frac{ \left( \frac{\Beta^2_1}{\Beta} \right)^2 \eta_1}{1 + \frac{\Beta^2_1}{\Beta} \eta_1}\right]\right)^{-1}, \\
\eta_2  &=& \left( \eta^{-2}_1 -  \alpha \mathbb{E} \left[ \left(\frac{\frac{\Beta^2_1}{\Beta}}{1 + \frac{\Beta^2_1}{\Beta} \eta_1} \right)^2\right] \right)^{-1}.
\end{eqnarray}

Similarly, to evaluate the expression for interference $P_{\mathsf{inter}}$, $\trace \left \{ \mathbf{Z} \right \}$ can be expanded as in equation~(\ref{eq:expandedTraceOfZ}), where, in step $(a)$  we use $\trace\{\mathbf{qq^H}\} = \mathbf{q^Hq}$ and if 
\setcounter{equation}{58}
\begin{eqnarray}
\mathbf{S}_k =  \sum_{k\neq 1,k} \nu_k \mathbf{H}_k\mathbf{a}_k \mathbf{a}^H_k\mathbf{H}^H_k  + (\theta_1 + \theta_2 +\sigma^2) \mathbf{I}  
\end{eqnarray} then in step $(b)$ use matrix inversion lemma as,
\begin{eqnarray}
\mathbf{S}^{-1} = \mathbf{S}^{-1}_k \left( \mathbf{I} - \frac{\nu_k\mathbf{H}_k \mathbf{a}_k \mathbf{a}^H_k \mathbf{H}^H_k \mathbf{S}^{-1}_k}{1 + \nu_k \mathbf{a}^H_k\mathbf{H}^H_k\mathbf{S}^{-1}_k \mathbf{H}_k\mathbf{a}_k}\right) \text{.}
\end{eqnarray}
Notice that using Lemma~\ref{lemma:TraceLargeMatrix}, the terms  $\mathbf{H}^H_k \mathbf{S}^{-1}_k \mathbf{H}_k \overset{a.s.}{\longrightarrow} \eta_1 \mathbf{I}$ and $\mathbf{H}^H_k \mathbf{S}^{-2}_k \mathbf{H}_k \overset{a.s.}{\longrightarrow} \eta_2 \mathbf{I}$ and it appears repeatedly in equation~(\ref{eq:expandedTraceOfZ}). Therefore, in the limit of infinite number of antennas, and for a given $\alpha$, using Lemma \ref{lemma:TraceLargeMatrix} we have 
\begin{eqnarray}
\mathbf{H}^H_1 \mathbf{Z} \mathbf{H}_1  \overset{a.s.}{\longrightarrow} \frac{ \trace\{ \mathbf{Z}\}}{M}  \mathbf{I}
\end{eqnarray}
and $ \trace\{ \mathbf{Z}\}/M $ in turn converges to the expression in equation~(\ref{eq:TraceInterMatrix}). In equation~(\ref{eq:TraceInterMatrixLastbut1}) the third term $\mathbb{E}\left[ \frac{\frac{\Beta^2_1}{\Beta} \eta_2}{\left( 1 + \frac{\Beta^2_1}{\Beta} \eta_1\right)^2}\right]$ is equal to  $\int^{\infty}_{0} \frac{\lambda}{( \lambda + \bar{\theta}_1  + \bar{\theta}_2+ \sigma^2)^2} dG(\lambda)$. This follows from corollary \ref{cor:JZhang}.
 
  Using equations~(\ref{eq:SignalPowerTrace}),~(\ref{eq:NoisePowerTrace}),~(\ref{eq:TraceInterMatrix}) in expressions (\ref{eq:SgnlPwrMmse}),~(\ref{eq:noisePwrMmse}),~(\ref{eq:pilotInterPwrMmse}) and (\ref{eq:InterPwrMmse}) the SINR given in equation~(\ref{eq:expressionSINR}) converges almost surely to $\sinrmmsePil$ as in expression~(\ref{eq:SINRMmseLargeSystem}). 
  
\section{Proof of Proposition \ref{th:SINRLargeSystem}}
\label{Appendix:ProofSINRMatchedLargeSys}
With $\nu_1 = \left( \frac{\beta_{11}}{\beta^{(1)}}\right)^2$, using the matched filter given by $\mathbf{c} = \frac{\sqrt{\beta_{11}}}{ \beta^{(1)}}\mathbf{H}_{1}\mathbf{a}_1$, the signal power, noise power, pilot interference power is given by $P_{\mathsf{signal}} = \nu_1 \left|\mathbf{a}_1\mathbf{H}^H_1\mathbf{H}_1 \mathbf{e}_1\right|^2$, $P_{\mathsf{noise}} = \frac{\nu_1}{\beta_{11}} \sigma^2 \mathbf{a}^H_1 \mathbf{H}^H_1 \mathbf{H}_1 \mathbf{a}_1$, $P_{\mathsf{contam}} = \frac{\nu_1}{\beta_{11}} \sum^B_{j=1}\beta_{j1} \left| \mathbf{a}^H_1 \mathbf{H}^H_1 \mathbf{H}_1 \mathbf{e}_j\right|^2$. Since, $\mathbf{H}^H \mathbf{H} \overset{a.s.}{\longrightarrow} \mathbf{I}$ as $M \rightarrow \infty$, we have, $P_{\mathsf{signal}} \overset{a.s.}{\longrightarrow} \nu_1 \beta_{11}$, $P_{\mathsf{noise}} \overset{a.s.}{\longrightarrow} \frac{\nu_1}{\beta_{11}} \beta^{(1)}\sigma^2$, $P_{\mathsf{contam}} \overset{a.s.}{\longrightarrow}                 \frac{\nu_1}{\beta_{11}} \sum^B_{j=2}\beta^2_{j1}$.
Using Lemma \ref{lemma:TraceLargeMatrix} in the interference term we have,
\setcounter{equation}{63}
\begin{eqnarray}
P_{\mathsf{inter}} &=& \frac{\nu_1}{\beta_{11}} \mathbf{a}^H_1 \mathbf{H}^H_1 \left( \sum^B_{j=1} \sum^K_{k=2} \beta_{jk} \mathbf{h}_k \mathbf{h}^H_k\right)\mathbf{H}_1 \mathbf{a}_1,  \nonumber \\
& \overset{a.s.}{\longrightarrow} & \frac{\nu_1}{\beta_{11}} \mathbf{a}^H_1 \left(\frac{1}{M} \trace \left\{ \sum^B_{j=1} \sum^K_{k=2} \beta_{jk} \mathbf{h}_k \mathbf{h}^H_k \right\} \mathbf{I} \right)  \mathbf{a}_1, \nonumber \\
&=& \frac{\nu_1}{\beta_{11}} \alpha \sum^B_{j=1} \beta_{j1}\frac{1}{K} \left( \sum^B_{j=1} \sum^K_{k=2} \beta_{jk} \mathbf{h}^H_k \mathbf{h}_k\right), \nonumber \\
&=& \frac{\nu_1}{\beta_{11}} \alpha \sum^B_{j=1} \beta_{j1}  \left( \sum^B_{j=1} \mathbb{E} [\Beta_j]\right).
\end{eqnarray}
Rearranging  the terms in equation~(\ref{eq:expressionSINR}) we get the expression for $\sinrmatPil$.

\bibliographystyle{IEEEbib}

\end{document}